\documentclass[10pt]{iopart}
\newif\ifpdf\ifx\pdfoutput\undefined\pdffalse\else\pdfoutput=1\pdftrue\fi
       \newcommand{\pdfgraphics}{\ifpdf\DeclareGraphicsExtensions{.pdf,.jpg}\else\fi}
\usepackage{graphicx}
\usepackage{hyperref}
\usepackage{amssymb}
\usepackage{longtable}
\usepackage{rotating}
\usepackage{color}
\usepackage{epsfig}
\usepackage{epsf}
\usepackage{fancyhdr}
\def\be{\begin{equation}}
\def\ee{\end{equation}}
\def\bea{\begin{eqnarray}}
\def\eea{\end{eqnarray}}
\def\M{{\cal M}}

\def\Tisco{T_{\textrm{\mbox{\tiny{ISCO}}}}}
\def\fisco{f_{\textrm{\mbox{\tiny{ISCO}}}}}
\def\qnr{{\textrm{\mbox{\tiny{QNR}}}}}
\def\ins{{\textrm{\mbox{\tiny{INS}}}}}
\def\rms{{\textrm{\mbox{\tiny{rms}}}}}
\def\rd{{\textrm{\mbox{\tiny{RD}}}}}
\def\gl{{\textrm{\mbox{\tiny{GL}}}}}

\def\Tqnr{T_{\textrm{\mbox{\tiny{QNR}}}}}

\begin{document}
\pdfgraphics
\title[Parameter estimation of compact binaries]
{Parameter estimation of compact binaries using the inspiral and
ringdown waveforms}

\author{Manuel Luna\dag\ and Alicia M. Sintes\dag\ddag}

\address{\dag\ Departament de F\'{\i}sica, Universitat de les Illes
Balears, Cra. Valldemossa Km. 7.5, E-07122 Palma de Mallorca,
Spain}
\address{\ddag\ Max-Planck-Institut f\"ur
    Gravitationsphysik, Albert Einstein Institut, Am M\"uhlenberg 1,
    D-14476 Golm, Germany}

\address{E-mail: \texttt{manuel.luna@uib.es}, \texttt{sintes@aei.mpg.de} }

\begin{abstract}
 We analyze the problem of parameter estimation for compact binary systems
 that could be detected by ground-based gravitational wave detectors.
 So far this problem has only been dealt with for the inspiral and 
 the ringdown phases separately. In this paper, we combine the information 
 from both signals, and we study the improvement in parameter estimation, at a
 fixed signal-to-noise ratio, 
 by including the ringdown signal without making any assumption
 on the merger phase. The study is performed for both initial and advanced
 LIGO and VIRGO detectors.
 
\end{abstract}

\date{\today}

%

\section{Introduction}
\label{sec:intro}

Coalescing compact binaries consisting of either black holes (BH)
or neutron stars (NS)
are among the targets of on-going searches for gravitational waves in 
the data of ground-based interferometric detectors
such as GEO\,600 \cite{GEO1,GEO2}, the Laser Interferometer
Gravitational Wave Observatory (LIGO) \cite{ligo1, ligo2}, TAMA\,300
\cite{tama95} and VIRGO \cite{virgo97}. 

The coalescence of a compact binary
system is commonly divided into three stages, which are not
very well delimited one from another, namely the
inspiral, the merger and the ringdown. 
Many studies so far have focused on the gravitational waves emitted
during the inspiral phase  because the inspiral waveform
is very well understood 
\cite{lucLR, luc95,luc95L,luc96cqg,damour98,damour00,luc04L,luc05} 
and the event rates seem promising
\cite{Phinney,vicky05,vicky05b,vicky02,vicky01}. Gravitational waves from the merger
can only be calculated using the full Einstein equations. Because of the
extreme strong field nature of this epoch neither a straightforward application of
post-Newtonian theory nor perturbation theory is very useful. Recent numerical 
work \cite{bruegmann04,pretorius05,baker01,baker02a, baker02b} has given some insights into the merger 
problem, but  there are no
reliable models for the waveform of the merger phase at this time.
The gravitational radiation from the ringdown phase is also well known
and it can be described by quasi-normal modes \cite{Echevarria:1988}.
In spite of the importance of the ringdown there are a fewer publications
on ringdown searches compared to those for inspiral searches.

Flanagan and Hughes \cite{Flanagan:1997sx} were the first in studying
the contribution of the three phases to the signal-to-noise ratios 
both for ground-based and also space-based
interferometers, but they did not study the problem of accuracy in the 
parameter estimation. This is an important problem because
many efforts  are now underway to detect both the inspiral and the ringdown
signals using matched filtering techniques in real data 
\cite{Tsunesada,Nakano,ligoS2a,ligoS2b,ligo}.
In a recent paper \cite{Arun:2004hn},
parameter estimation of inspiralling compact binaries has been revised using 
up to the 3.5 restricted post-Newtonian approximation,
 extending previous analysis \cite{Cutler:1994ys,Poisson95}, 
  but ignoring the other stages.
The parameter estimation for the ringdown phase alone has also been studied,
 some time ago, for ground based detectors 
 \cite{Finn:1992wt,Creighton99,japon} as well as for 
 LISA \cite{0512160WillLISA}.
The aim of this paper is to discuss how parameter estimation can be improved
by using information from both the inspiral and the ringdown phases combined
together in  matched filtering like analysis for different ground-based
detectors.

This paper is organized as follows: Section~\ref{sec:parameter}  introduces
our notation and reviews the basic concepts of signal parameter estimation in
matched filtering. Section~\ref{sec:noise} provides the noise curves used in
this study for initial and advanced LIGO and VIRGO.
Section \ref{sec:waveform} briefly describes the waveforms that we
are looking for.  For the inspiral phase, we consider a non-spinning compact
system with circular orbits and the waveform in the restricted post-Newtonian
approximation. For the ringdown, we assume that the dominant mode has $l=m=2$
and therefore the waveform is given by an exponentially decaying sinusoid.
Section~\ref{sec:pe} studies the impact on the parameter estimation for
coalescing binary black holes, by combining
the signals from both the inspiral and ringdown phases and compares the results
with the case of inspiral phase alone. The results are presented for a fixed
inspiral signal-to-noise ratio of 10. Different number of parameters are used
as well as different values for the ringdown efficiency. Finally 
section~\ref{sec:conc} concludes with a summary of our
results and plans for further work. 
 In the Appendices we collect various technical calculations
and we present an explicit analytical calculation of the Fisher matrix
for the ringdown phase that has been used to compare with 
the numerical results.


\section{Summary of parameter estimation}
\label{sec:parameter}

In this section we briefly review the basic concepts and formulas of
signal  parameter estimation relevant to the goal of this paper;
we refer the reader to \cite{Cutler:1994ys} for a more detailed analysis.

The output of a gravitational wave detector can be schematically 
represented as
\be
s(t) = h(t) + n(t) \, ,
\ee
where  $n(t)$ is the noise that affects the observation and $h(t)$ is the
gravitational wave signal measured at the detector,  a linear
superposition of the two independent polarizations of the strain
amplitude $h_+$ and $h_\times$, given by
\begin{equation}
h(t) = F_+(\theta, \phi,\psi)h_+(t) + F_\times(\theta, \phi,\psi)h_\times(t) 
\, ,
\end{equation}
where $F_+$ and $F_\times$ are the antenna 
pattern functions, that
depend on the direction of the source in the sky $(\theta, \phi)$ 
and the polarization angle $\psi$. 
In case of a laser interferometer detector, the expressions of 
$F_+$ and $F_\times$ are given by \cite{Th300}:
\bea
F_+(\theta,\phi,\psi)&=&{1\over 2}(1+\cos^2\theta)\cos 2\phi \cos 2\psi
-\cos\theta\sin 2\phi \sin 2\psi \ , \\
F_{\times}(\theta,\phi,\psi)&=&{1\over 2}(1+\cos^2\theta)\cos 2\phi \sin 2\psi
+\cos\theta\sin 2\phi \cos 2\psi \ .
\eea

For sake of simplicity we shall made the standard
assumptions that the noise $n(t)$ has zero mean and it 
is stationary and Gaussian, although in
realistic cases this hypothesis is likely to be violated at some level.
Within this approximation, the Fourier components 
of the noise
are statistically described by:
\be
E[\tilde{n}(f)\tilde{n}^*(f')] =  \frac{1}{2} \delta (f-f')S_n(f) \, ,
\ee
where $E[]$ denotes the expectation value with respect to an ensemble of noise
realization, the $*$ superscript denotes complex conjugate,
$S_n(f)$ is the one sided  noise power spectral density, and 
tildes denote Fourier transforms according to the convention
\be
\label{fourier_transform}
\tilde{x}(f) = \int_{-\infty}^{\infty}  e^{i2 \pi f t} x(t)\;  dt \, .
\ee
%

With a given noise spectral density for the detector, one defines
the ``inner product'' between any 
two signals $g(t)$ and $h(t)$ by:
\begin{equation}\label{inner_product}
( g | h ) \equiv 2 \int_0^{\infty} \frac{\tilde{g}^*(f) \tilde{h}(f)
+ \tilde{g}(f) \tilde{h}^*(f)}{S_n (f)} \; df \, .
\end{equation}
With this definition, the probability of the noise to have a
realization $n_0$ is just:
\be
p(n=n_0) \propto e^{(n_0 | n_0)/2} \, .
\ee

The optimal signal-to-noise ratio (SNR) $\rho$, achievable with linear methods
({\it e.g.}, matched filtering the data) is given by the standard expression
\begin{equation}
\label{snr_2}
\rho^2=( h | h ) = 4 \int_0^{\infty} \frac{|\tilde{h}(f)|^2}{S_n(f)}\; df \, .
\end{equation}

In the limit of large SNR, if the noise is stationary and Gaussian, the 
probability that the gravitational wave signal $h(t)$ is characterized by a
given set of values of the source parameters 
$\mbox{\boldmath $\lambda$}=\{\lambda^k\}_k$ is given by a Gaussian
probability  of the form \cite{Finn:1992wt}:
\be
\label{p10}
p( \mbox{\boldmath $\lambda$} \vert h)=p^{(0)}(\mbox{\boldmath $\lambda$})
 \exp \left[ - \frac{1}{2} \Gamma_{j k} \Delta\lambda^j \Delta\lambda^k \right] \,
 \, ,
\ee
where $\Delta\lambda^k$ is the difference between the true value of the
parameter and the best-fit parameter in the presence of some realization of the
noise, 
$p^{(0)}(\mbox{\boldmath $\lambda$})$ represents the distribution of prior
information (a normalization constant) and 
 $\Gamma_{j k}$ is the so-called Fisher information matrix defined by
\begin{equation}
\label{fisher_matrix}
\Gamma_{i j} \equiv (\partial_i h | \partial_j h)
= 2 \int_0^{\infty} \frac{\partial_i \tilde{h}^*(f)
\partial_j \tilde{h}(f)
+ \partial_i \tilde{h}(f) \partial_j \tilde{h}^*(f)}{S_n (f)} \; df \, ,
\end{equation}
where $\partial_i = \frac{\partial}{\partial \lambda^i}$.

The inverse of the Fisher matrix, known as the 
variance-covariance matrix, gives us the accuracy with 
which we expect to measure
the parameters $\lambda^k$
\begin{equation}
\label{covariance_matrix}
\Sigma^{jk} \equiv 
(\Gamma^{-1})^{jk} = \langle \Delta\lambda^j\Delta\lambda^k\rangle \, .
\end{equation}
Here the angle brackets denote an  average over the probability
distribution function in Eq.~(\ref{p10}).
The root-mean-square error $\sigma_k$ in the estimation of the parameters
$\lambda^k$ can then be calculated, in the limit of large SNR, by taking the
square root of the diagonal elements of the variance-covariance matrix,
\be
\sigma_k = \langle (\Delta\lambda^k)^2\rangle^{1/2} =\sqrt{\Sigma^{kk}} \, ,
\ee
and the correlation coefficients  $c^{jk}$ between two 
parameters $\lambda^j$ and
$\lambda^k$ are given by:
\be
c^{jk}= \frac{\langle \Delta\lambda^j\Delta\lambda^k\rangle}{\sigma_j\sigma_k}=
\frac{\Sigma^{jk}}{\sqrt{\Sigma^{jj}\Sigma^{kk}}} \, .
\ee

\section{Noise spectra of the interferometers}
\label{sec:noise}
In this paper, we use three different noise curves to understand the effect of
detector characteristics on the parameter estimation.
The noise curves used are initial and advanced LIGO and VIRGO as in
\cite{Arun:2004hn}. Those are:

For the initial LIGO
\be
\label{noise_initial_ligo}
S_n (f) = \left\{ \begin{array}{ll}
S_0 \, [(4.49 x)^{-56} + 0.16 x^{-4.52} + 0.52 + 0.32 x^2]\, , & f\ge f_s\\
 \infty \; , & f < f_s
 \end{array}
\right.
\ee
where $x=f/f_0$, with $f_0 = 150$~Hz (a scaling frequency chosen for
convenience), $f_s=40$~Hz is the lower cutoff frequency, and 
$S_0=9 \times 10^{-46}$~Hz$^{-1}$.

For advanced LIGO the noise curve is given by
\be
\label{noise_advance_ligo}
S_n (f) = \left\{ \begin{array}{ll}
 S_0 \, \Bigg[ x^{-4.14} - 5 x^{-2} + \frac{111(1-x^2+x^4/2)}
{(1+x^2/2)} \Bigg] \; , & f \ge f_s\\
\infty \; , & f < f_s 
\end{array}\right.
\ee
where 
$f_0=215$~Hz, $f_s=10$~Hz
and $S_0=10^{-49}$~Hz$^{-1}$.

Finally, for the VIRGO detector the expected noise curve is given by:
\be
\label{noise_virgo}
S_n (f)  = \left\{ \begin{array}{ll}
S_0 \; [(6.35x)^{-5} + 2x^{-1} + 1 + x^2] \; , & f \ge f_s\\
 \infty \; , & f < f_s
 \end{array}\right.
\ee
where $f_0=500$~Hz, $f_s=20$~Hz
 and $S_0 = 3.24 \times 10^{-46}$~Hz$^{-1}$.

\section{The gravitational-wave signal}
\label{sec:waveform}

As discussed in the introduction, the coalescence and its associate
gravitational wave signal can be divided into three successive epochs in the
time domain: inspiral, merger and ringdown. During the inspiral the distance
between the stars diminishes and the orbital frequency sweeps up. For low-mass
binary systems, the waveforms are well modeled using the post-Newtonian
approximation to general relativity \cite{lucLR,luc95L,luc96cqg,luc04L}. 
Eventually the post-Newtonian
description of the orbit breaks down, and the black holes cannot be
treated as point particles any more. What is more, it is expected that they
will reach the \emph{innermost stable circular orbit} (ISCO), at which the
gradual inspiral ends and the black holes begin to plunge together to
 form a single black hole.
This is referred as the merger phase. At present,  the merger phase
 is not well understood and no analytical reliable waveforms exist. At the end,
 the final black hole will gradually settle down into a Kerr black hole.
The last gravitational waves 
 will be dominated by the 
 quasi-normal ringing modes of the black hole  (see \cite{qnmLR} and references
 therein) and can be treated using perturbation theory \cite{leaver85}.
 At late time, the radiation will be dominated 
 by the $l=m=2$ mode \cite{Echevarria:1988}. 
 This is  the so-called ringdown phase.

The gravitational waveform of coalescing compact binaries thus takes the form
\begin{equation}
\label{wform_global_1}
h(t)=\left\{ \begin{array}{lc}
h_{\textrm{\mbox{\tiny{inspiral}}}}(t) & -\infty< t< \Tisco \\
h_{\textrm{\mbox{\tiny{merger}}}}(t) & \Tisco<t<\Tqnr \\
h_{\textrm{\mbox{\tiny{ringdown}}}}(t) & \Tqnr<t < \infty 
\end{array} \right.
\end{equation}
where $\Tisco$ is the time when the system reaches the ISCO and 
$\Tqnr$ is the time when the quasi-normal mode  $l=m=2$ begins to dominate the
ringdown, although there is some arbitrariness in choosing $\Tisco$ and
$\Tqnr$ to delimit  the three epochs. 


\subsection{The inspiral waveform}
\label{ss:inspiral}

For a non-spinning compact binary system with circular orbits,
the two polarizations  $h_+$ and $h_{\times}$ of the inspiral waveform
can be well described by the 
post-Newtonian expansion. Thus setting $G=c=1$, they read:
\bea
\label{expansion}
h_{+,\times}= {2M\eta\over r}(M\omega)^{2/3} & &\left\{H_{+,\times}^{(0)}
+v^{1/2}H_{+,\times}^{(1/2)}+v H_{+,\times}^{(1)}+ \right.\nonumber\\ 
& & \left.
\,\, +v^{3/2}H_{+,\times}^{(3/2)}
+v^2 H_{+,\times}^{(2)}+ \cdots \right\} \ ,
\eea
where $v\equiv (M\omega)^{2/3}$, $\omega$ is the  orbital frequency, 
$r$ is the distance to the source, $M=m_1+m_2$ is the total mass,
$\mu=m_1m_2/M$ is the reduced mass, $\eta=\mu/M$ is the symmetric mass
ratio and ${\cal M}=\mu^{3/5}M^{2/5}=M\eta^{3/5}$ is the chirp mass.

In what follows  we consider the waveform in the restricted post-Newtonian 
approximation 
\cite{Sintes:1999cg},
corresponding to a frequency twice the orbital frequency, and we ignore
higher order harmonics. This corresponds to the lowest terms in the series
(\ref{expansion}). The functions $H_+^{(0)}$,  $H_{\times}^{(0)}$
are given by \cite{luc96cqg}:
\bea
H_+^{(0)}&=& -(1+\cos \iota^2)\cos \Phi(t) \ , \\
H_{\times}^{(0)}&=& -2\cos \iota \, \sin \Phi(t) \ , 
\eea
with  $\iota$ being the angle between the orbital angular
momentum of the binary and the line of sight from the detector to the source.
 $\Phi$ is the phase of the gravitational wave an the instant $t$, that we 
 consider modeled through 2nd post-Newtonian order, neglecting 
 the higher order terms in this analysis, since they would not contribute
 significantly to the result.

The Fourier transform of the inspiral waveform can be computed using 
 the stationary phase approximation
\cite{Cutler:1994ys,Poisson95,Sa91,Fi93}
 and this yields:
\be
\label{tform_inspiral}
\tilde{h}_\ins(f)= \left\{ \begin{array}{lc}
\mathcal{A}_\ins f^{-7/6}e^{i \Psi (f)} & f<\fisco\\
0 & f>\fisco
\end{array} \right. \, ,
\ee
with 
\be
\label{amplitude_inspiral}
\mathcal{A}_\ins= -\frac{\M^{5/6}}{r}
\sqrt{{5\pi\over 96}}\pi^{-7/6}
\sqrt{F_+^2(1+c^2)^2+F_{\times}^24c^2} \, ,
\ee
\begin{equation}
\label{inspiral_phase}
\Psi (f)=2 \pi f t_c - \phi_c - \frac{\pi}{4}+\frac{3}{128}
\sum_{k=0}^{4} A_k u^{k-5} \, ,
\end{equation}
 where $c=\cos \iota$, $t_c$ refers to the coalescence time, 
 $\phi_c$ is the phase
 at the coalescence instant,  $u=(\pi \mathcal{M} f)^{1/3}$, and
 the coefficients $A_k$ are given by
\begin{eqnarray}
\label{alpha_parameters}
A_0&=&1\\
A_1&=&0\\
A_2&=&\frac{20}{9}(\frac{743}{336}+\frac{11}{4} \eta)\eta^{-2/5}\\
A_3&=&-16\pi\eta^{-3/5}\\
A_4&=&10(\frac{3058673}{1016064}+\frac{5429}{1008}\eta+\frac{617}{144}\eta^2)
\eta^{-4/5} \label{alpha_parameters2}
\end{eqnarray} 
 We also consider the ISCO   to take place at a separation of $6M$,
  corresponding to a final frequency
  \be
  \fisco = \frac{1}{6^{3/2}\pi M} \, .
  \ee

\subsection{The ringdown waveform}
\label{ss:rd}

The ringdown portion of the gravitational wave signal we consider
can be described as 
the $l=m=2$ quasi-normal mode. Therefore the gravitational radiation 
in the time domain is expected as the superposition of two different 
damped sinusoids, although one of these exponentials could be invisible
in the actual waveform as discussed in~\cite{0512160WillLISA}.
In our study, we  assume that the ringdown waveform can be written 
as in~\cite{Echevarria:1988}, corresponding to a circularly polarized
wave. In this way we have
\bea
\label{gw_ringdown}
h_+ (t) &-& i h_\times ( t)= {A M \over r} {_2}S_2^2 (\iota, \beta, a)
  \nonumber\\
& & \times \exp \left[-i 2 \pi f_\qnr(t-t_0) 
-{\pi f_\qnr\over Q}(t-t_0)+i \varphi_0\right] \, ,
\eea
where $t_0$ is the start time of the ringdown, $\varphi_0$ the initial phase,
 $M$ is the total mass of the system mass (see \cite{Hughes:2004vw} for further
 discussions),
$f_\qnr$ and $Q$ are the central frequency and the quality factor of the
ringing. For this mode, a good fit to the frequency $f_\qnr$ and 
quality factor $Q$, within an accuracy of  5\%, is
\begin{eqnarray}
\label{f_qnr}
f_\qnr &\approx& [ 1- 0.63(1-a)^{3/10} ] \frac{1}{2 \pi M}\, ,\\ 
\label{Q}
Q &\approx& 2 (1-a)^{-9/20} \, ,
\end{eqnarray}
where $aM^2$ is its spin, and $a$ is the Kerr parameter that lies 
in the range $(0.0, 0.998)$.
In our study we set $a$ to the near extremal value of 0.98 
(as in \cite{Flanagan:1997sx}),
although we consider $a$ as any other independent parameter 
when evaluating the Fisher matrix.
The function ${_2}S_2^2$ is the spin weighted spheroidal
harmonic that depends on the inclination angle of the black hole axis
seen from the observer and the Kerr
parameter $a$. 
%
$A$ is a dimensionless coefficient describing the magnitude of the
 perturbation when the ringdown begins. Although the value of the amplitude 
 is uncertain, we set the amplitude of this mode
by assuming that a fraction
$\epsilon$ of the system's mass is converted into gravitational waves
 during the ringdown~\cite{Flanagan:1997sx}
\begin{equation}
\label{energy_radiated_ringdown}
E_\rd \approx \frac{1}{8} A^2 M^2 f_\qnr Q  =
 \epsilon M\left( \frac{4 \mu}{M}\right)^2 \, .
\end{equation}
Therefore
\be
\label{intrinsic_amplitude}
A=\sqrt{\frac{128 \, \eta^2 \, \epsilon}{M f_{\qnr} Q}} \, .
\ee
The strain  produced at the detector can be written as:
\begin{equation}
\label{waveform_ringdown}
h_\rd(t)= \mathcal{A}_\rd \; \exp\left[-{(t-t_0) \pi f_\qnr\over Q}\right]
\;\cos( -2 \pi f_\qnr (t-t_0) + \gamma_0)
\end{equation}
where 
\begin{equation}
\label{amplitude_ringdown}
\mathcal{A}_\rd \equiv {A M \over
 r}
\sqrt{F_+ ^2 + F_\times^2} \; |{_2}S_2^2| \, .
\end{equation}
The Fourier transform of the waveform becomes: 
\bea
\label{tform_ringdown}
\tilde{h}_\rd(f)&=& \frac{\mathcal{A}_\rd}{2 \pi} ~e^{i 2 \pi f t_0} \\
& & \times \left(
\frac{e^{i \gamma_0}}{{f_\qnr \over Q} -2 i(f-f_\qnr)}
+\frac{e^{-i \gamma_0}}{{f_\qnr \over Q} -2 i(f+f_\qnr)}
\right)  \, . \nonumber
\eea

\section{Parameter estimation  of compact binaries  using 
the inspiral and  ringdown waveforms}
\label{sec:pe}

In this paper we want to study the impact on the parameter extraction
by  combining the signals from the inspiral and the ringdown epochs,
 neglecting all information coming from the merger epoch itself since no
 reliable  waveforms exist so far.

Following earlier works, we choose the set of independent parameter 
$\lambda_\ins$ describing the inspiral signal  to be
\be
\label{parameter_inspiral}
\lambda_\ins=
\{ \ln \mathcal{A}_\ins, f_0 t_c, \phi_c, \ln \mathcal{M}, \ln \eta \} \, ,
\ee
while for the ringdown, the parameters could be
\be
\label{parameter_ringdown_0}
\lambda_\rd=
\{ \ln \mathcal{A}_\rd, \ln M, \ln a, \gamma_0, t_0 \} \, .
\ee

A possible  approach to this  problem of parameter extraction  would be to
consider the two sets  of parameters (\ref{parameter_inspiral}) and
(\ref{parameter_ringdown_0}) as independent, perform matched filtering 
using the two template families (for the inspiral and the ringdown waveforms)
and then reduce the uncertainties in the parameter estimation
(in particular for the masses)  
by making a posterior consistency check~\cite{sathyaP}. 
However  in this paper we follow a different approach. We consider only a
single coalescing waveform, as if we were performing
matched filtering with a single template family bank, 
as given by Eq.~(\ref{wform_global_1}),
which describes the different phases, but ignoring the information
from the merger phase,  and we focus our 
attention on how parameter estimation of the  $\lambda_\ins$ parameters can be
improved as compared to the case in which the inspiral waveform is used alone.
For this reason the study  presented here focuses  only in 
those mass ranges for which the inspiral  signal  alone could 
be detectable by the detector.
 This corresponds to a total mass of approximately
  1--100~$M_\odot$ for initial LIGO, 
1--400~$M_\odot$ for advanced LIGO,  and
1--200~$M_\odot$ for VIRGO. 

The global waveform considered here becomes in the Fourier domain
\be
\label{tform_global}
\tilde{h}_\gl (f)=\tilde{h}_\ins (f) + \tilde{h}_\rd (f) \, ,
\ee
where $\tilde{h}_\ins (f)$ and $\tilde{h}_\rd (f)$ are given by
equations (\ref{tform_inspiral}) and (\ref{tform_ringdown}) respectively.
This global waveform is completely determined by a
set of independent parameters, given by
\be
\label{global_parameter_set}
\lambda_\gl=\{ \ln \mathcal{A}_\ins, f_0 t_c, \phi_c, \ln \mathcal{M},
\ln \eta ,\ln a, \gamma_0, t_0 \} \, .
\ee
Notice that we do not include $\ln \mathcal{A}_\rd$ as an independent parameter
since for a given source location and orientation, and a given 
ringdown efficiency $\epsilon$, the ringdown amplitude $\mathcal{A}_\rd$,
given by Eq.~(\ref{amplitude_ringdown}), 
is determined by the inspiral amplitude $\mathcal{A}_\ins$, the Kerr parameter
and the masses. Instead, what we do is to find a heuristic relation between
$\mathcal{A}_\rd$ and $\mathcal{A}_\ins$ by averaging over source directions
and black-hole orientations, making use of the angle averages:
$\langle F_+^2\rangle_{\theta, \phi,\psi}
=\langle F_\times^2\rangle_{\theta, \phi,\psi} =1/5$, 
$\langle F_+ F_\times\rangle_{\theta, \phi,\psi} =0$, 
$\langle c^2\rangle_{\iota}= 1/3$,
$\langle (1+c^2)^2\rangle_{\iota}= 28/15$,
and 
$\langle |{_2}S_2^2|^2\rangle_{\iota, \beta} = 1/4 \pi$. The angle averaged 
root mean square  (rms) values of the
inspiral and ringdown amplitudes $\mathcal{A}_\ins$, $\mathcal{A}_\rd$,
given by Eq.~(\ref{amplitude_inspiral}) and (\ref{amplitude_ringdown}), become
\bea
\label{rms_amplitude_inspiral}
\mathcal{A}_\ins^\rms\equiv \sqrt{\langle \mathcal{A}_\ins^2\rangle}
 = \frac{1}{\sqrt{30}\pi^{2/3}}\frac{\mathcal{M}^{5/6}}{r}\\
\label{rms_amplitude_ringdown}
\mathcal{A}_\rd^\rms\equiv \sqrt{\langle \mathcal{A}_\rd^2\rangle}
= \frac{1}{\sqrt{10 \pi}} \frac{A M}{r} \, .
\eea

From the above equations, together with Eq.~(\ref{intrinsic_amplitude}), 
we derive a relation 
$\mathcal{A}_\rd=
\mathcal{A}_\rd(\mathcal{A}_\ins, \mathcal{M}, \eta, a)$ through
the ratio of rms of both amplitudes:
\be\label{amplitudes_relation}
\mathcal{A}_\rd(\mathcal{A}_\ins, \mathcal{M},\eta, a)
\equiv \frac{\mathcal{A}_\rd^\rms}{\mathcal{A}_\ins^\rms}\, \mathcal{A}_\ins
=\sqrt{\frac{384  \pi^{1/3} \epsilon}{M f_\qnr \, Q}} \eta^{2/5}
\mathcal{M}^{1/6} \, \mathcal{A}_\ins \, .
\ee
Note that in this relation the product of $M$ and $f_\qnr$
is just a function of $a$ as can be seen from Eq.(\ref{f_qnr}).
This relation (\ref{amplitudes_relation}) will be used in calculating the SNR
as well as the Fisher matrix for the global waveform.

\begin{figure}
  \begin{center}
  \begin{tabular}{c}
  (a) initial LIGO\\
   \includegraphics[height=6cm]{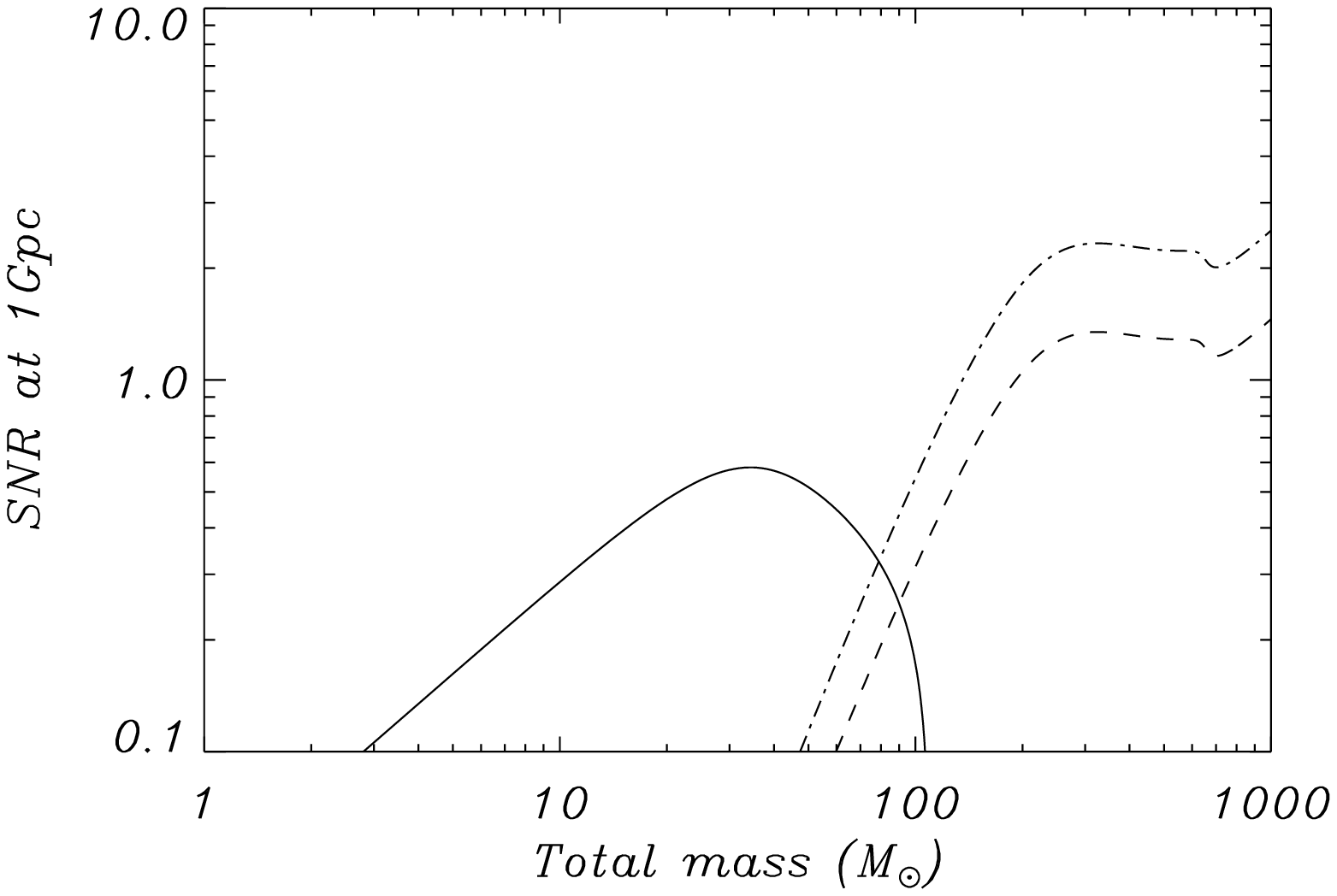} \\
   (b) advanced LIGO\\
   \includegraphics[height=6cm]{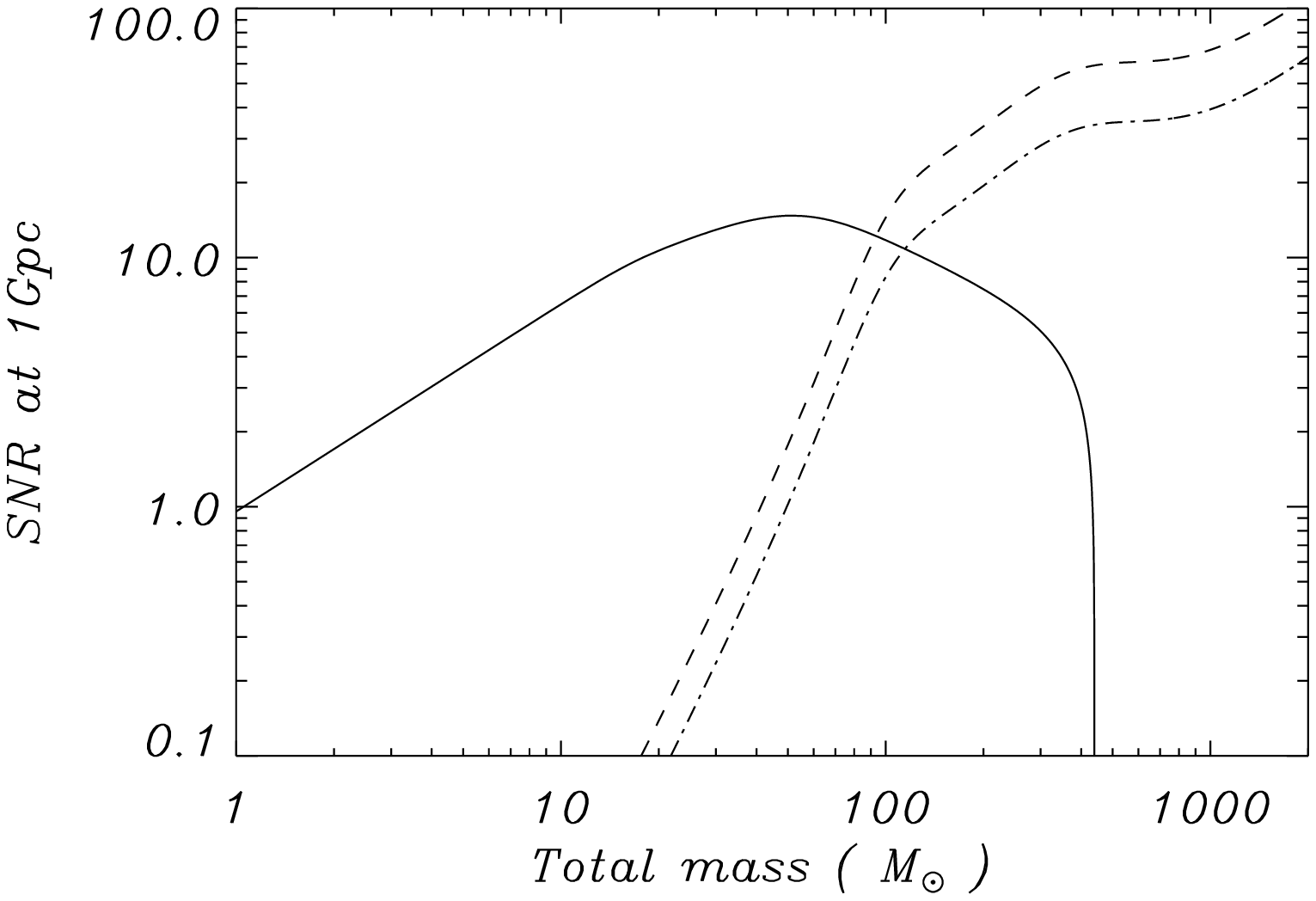} \\
   (c) VIRGO\\
   \includegraphics[height=6cm]{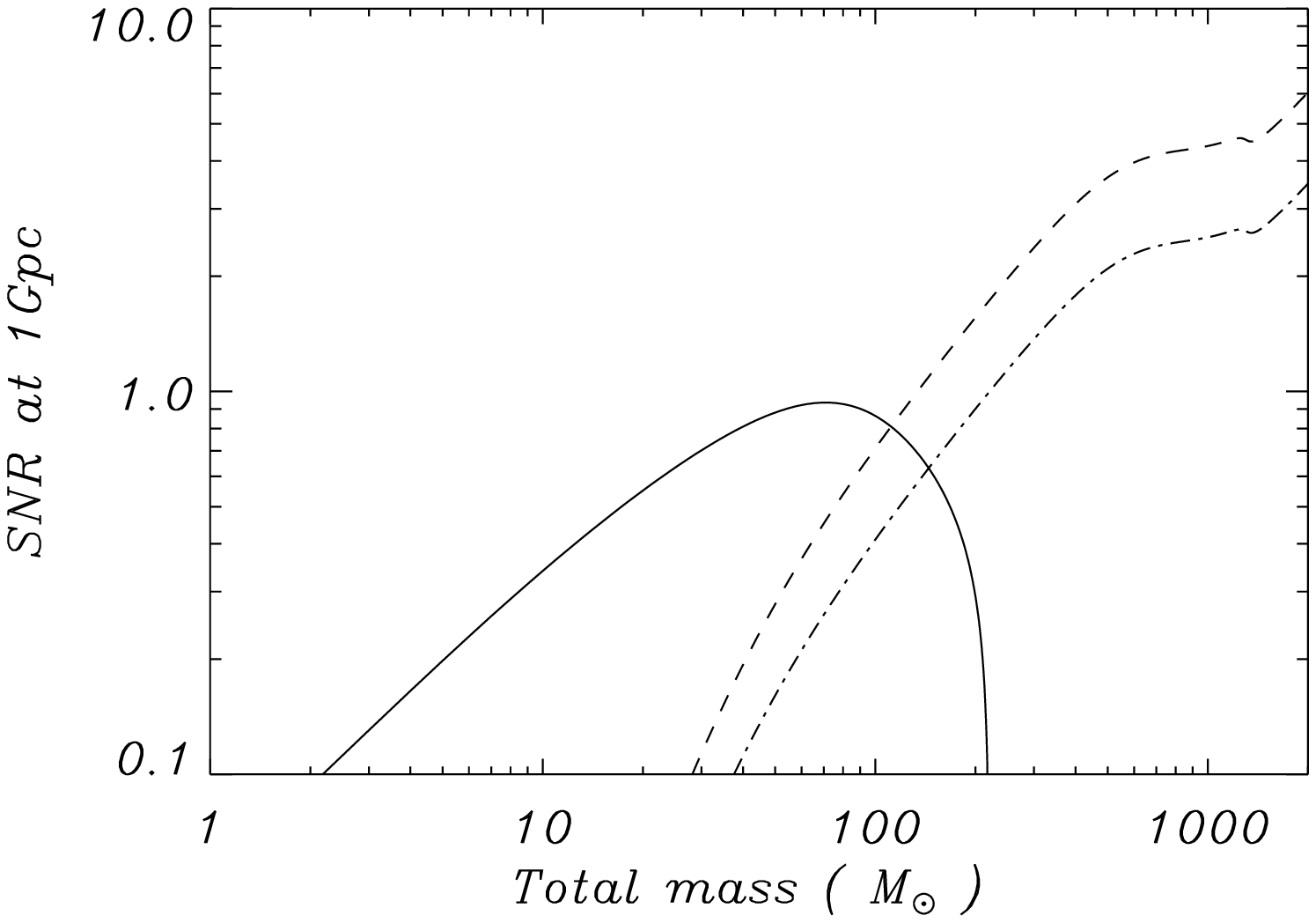} \\
  \end{tabular}
  \end{center}
  \caption{The  averaged signal-to-noise ratio 
  for equal-mass black hole
  coalescences detected by ground-based interferometers at a luminosity 
  distance of 1 Gpc. The solid line is the SNR curve for the inspiral,
  and the dash and dash-dotted lines for the ringdown portion of the signal 
  assuming a value of $a=0.98$ and
  $\epsilon$ equals to 1.5\% and 0.5\%, respectively. 
 The top panel corresponds to initial LIGO, the middle panel
  to advanced LIGO and the bottom one to VIRGO.
   }
  \label{f1}
\end{figure}

The SNR values for equal-mass  black hole binaries are shown in
 figure~\ref{f1}. This figure indicates that
 there is a range of masses (different for the different noise curves) 
 for which both the inspiral and the ringdown signals  could be
 detectable and one could search for both portions of the signal in order
 to improve the SNR and the accuracy in parameter 
 estimation\footnote{The calculation of the SNR for the 
 ringdown waveform is computed differently
from what was done by Flanagan and Hughes in~\cite{Flanagan:1997sx}. 
Instead of taking $\vert t-t_0\vert$ in the damped exponential, integrating
over $t$ over $-\infty$ to $+\infty$ and dividing the result by $\sqrt{2}$ to 
compensate for the doubling, we assume that the waveform
 $h_\rd (t)$  vanishes for $t<t_0$  and integrate only over $t>t_0$.}.


It is clear that both in the time domain, as well as, 
in the frequency domain, the
inspiral signal is decoupled from the ringdown one. The inspiral waveform
$\tilde{h}_\ins (f)$ ranges from the lower cut-off frequency $f_s$ to $\fisco$,
while the ringdown $\tilde{h}_\rd (f)$ is centered around $f_\qnr$ with a 
certain bandwidth, that in the literature is considered to be smaller than
$\Delta f /f_\qnr = 0.5$\footnote{Note that the distance between 
$f_\qnr$ and $\fisco$ is larger than the bandwidth of 
the ringdown signal $\tilde{h}_\rd (f)$. In particular if we consider the
value $a=0.98$ then $(f_\qnr-\fisco) / f_\qnr=0.833$ that suggests no overlap
between the inspiral and the ringdown signal.},
 although in our numerical simulations for 
the ringdown signal we use the same lower cut-off frequency $f_s$ 
and a higher cut-off frequency of 5000~Hz.
This justifies that the Fisher matrix, 
defined in  Eq.~(\ref{fisher_matrix}), of the global waveform
can be computed as
\be\label{elements_global_matrix}
\Gamma_{i j} = ( \partial_i h_\gl | \partial_j h_\gl )=
( \partial_i h_\ins | \partial_j h_\ins ) + ( \partial_i h_\rd |
\partial_j h_\rd ) \, ,
\ee
neglecting the  cross elements 
$( \partial_i h_\ins | \partial_j h_\rd )$. 
Therefore the Fisher matrix can be computed as the sum of Fisher matrix
of the inspiral waveform plus the Fisher matrix of the ringdown
\be
\label{global_matrix}
\Gamma_\gl = \Gamma_\ins + \Gamma_\rd \, ,
\ee
where we just need to be consistent in computing the elements 
corresponding to the same parameter set.
Also the total SNR is given by
\be
\label{global_snr}
\rho_\gl^2=\rho_\ins^2+\rho_\rd^2 \, .
\ee

The way we proceed is to analyze first the well known case 
of the inspiral signal alone, and then we compare the results with those
when using the inspiral and ringdown waveforms together. 
In order to separate the effects of increasing the number of parameters
from the fact we are using a more complex waveform, we study two different
cases:
\begin{itemize}
\item[i.] The case in which only the five inspiral parameters are considered.
This is equivalent to have no uncertainties in the spin of the final black hole,
nor in the initial phase and time of the ringdown signal. This of course,
would not be realistic in a  search, but it provides the optimal
improvement in parameter estimation one could expect from the fact that we added
the ringdown waveform.
\item[ii.] The more realistic case in which all the eight independent parameters 
given in~(\ref{global_parameter_set}) are considered.
\end{itemize}
In \ref{derivatives} the reader can find the explicit calculations
 of all the waveform derivatives necessary to compute the Fisher matrix
 which is then computed numerically.

In our analysis, we set the Kerr parameter $a=0.98$ 
(as in~\cite{Flanagan:1997sx}), and we consider two different 
values of $\epsilon$: a more optimistic value of 1.5\%,
(half the value used  in~\cite{Flanagan:1997sx}), and a more pessimistic one
of 0.5\%, that are more
consistent with recent numerical simulations \cite{baker01}. 
With these parameters we study  a 
range of masses, analyzing both the equal-mass  and unequal-mass cases,
for three ground-based detectors: initial LIGO, advanced
LIGO and VIRGO, using the noise curves described in section~\ref{sec:noise}. 
All the errors are computed at a fixed value of inspiral SNR of 10.

\begin{figure}
  \begin{center}
  \begin{tabular}{c}
  \includegraphics[height=6cm]{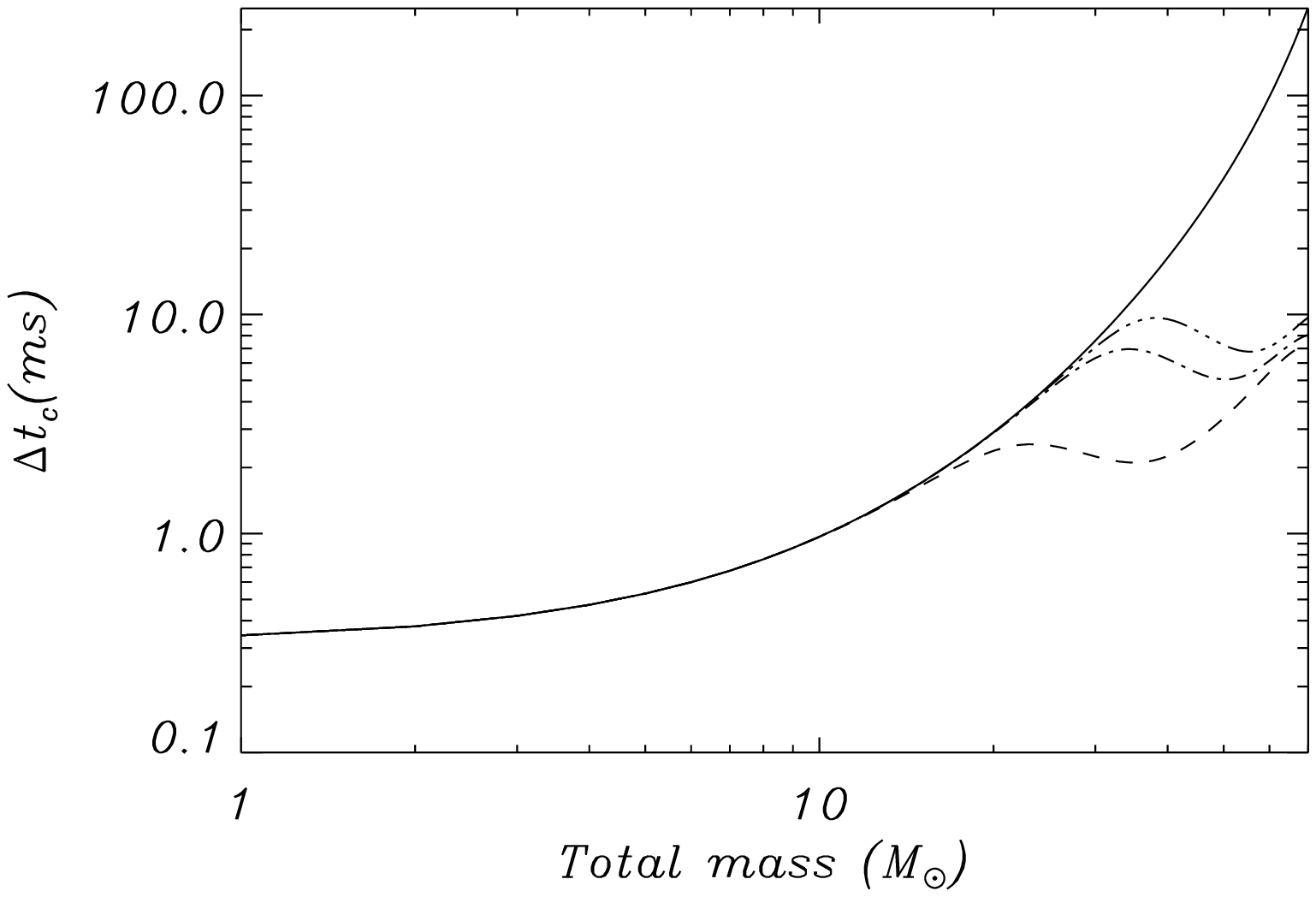}\\
  \includegraphics[height=6cm]{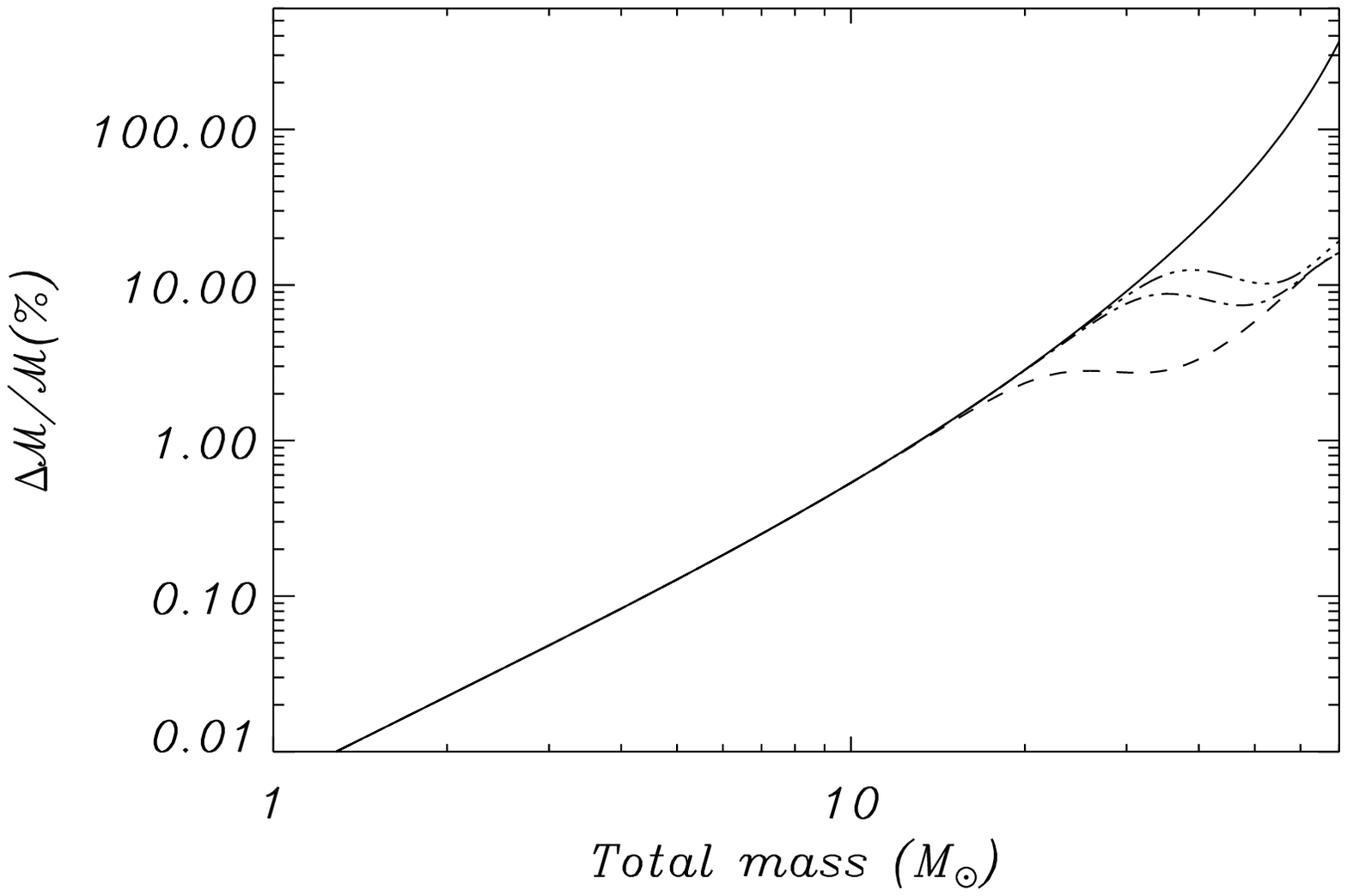} \\ 
  \includegraphics[height=6cm]{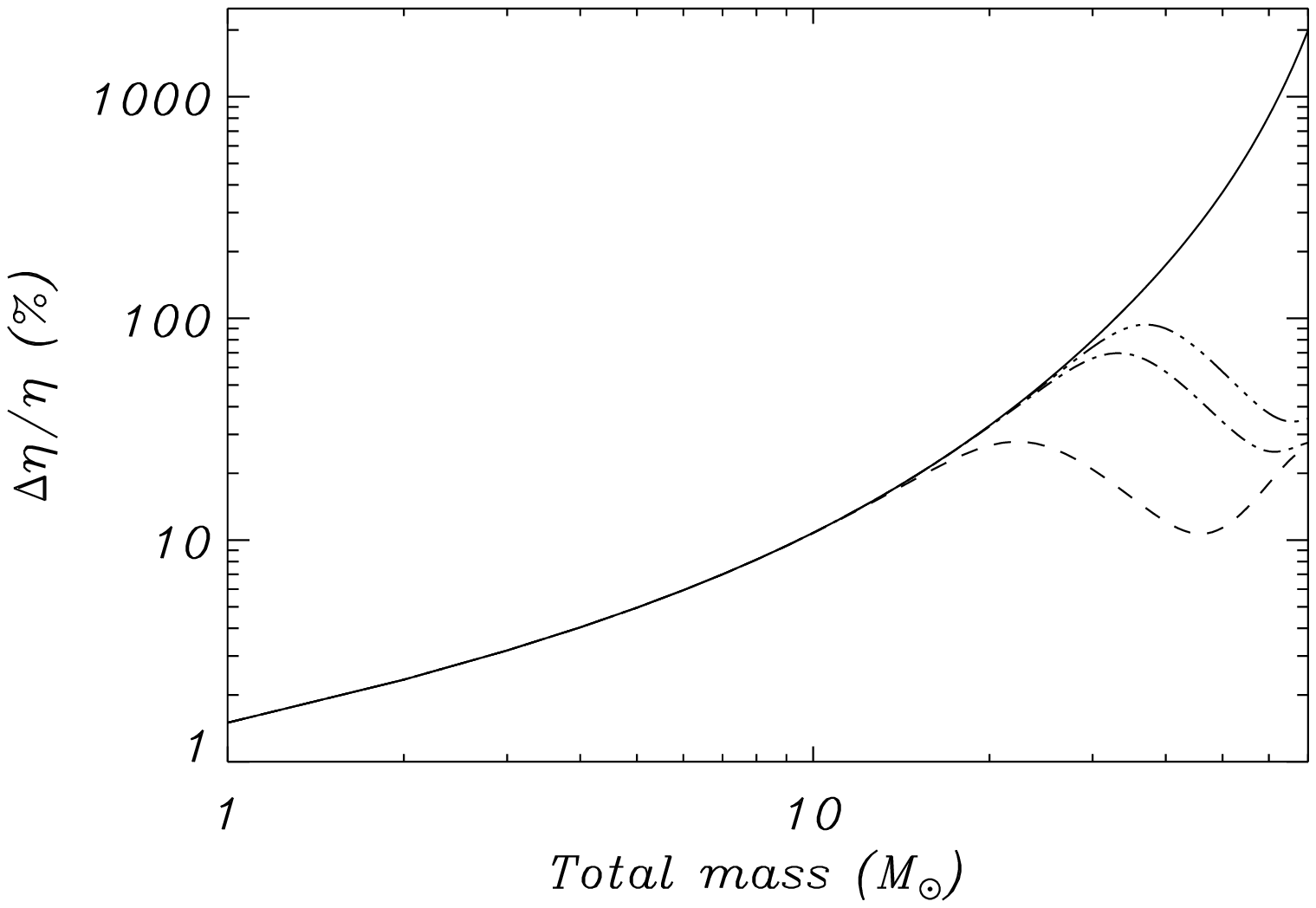}\\
  \end{tabular}
  \end{center}
  \caption{
  In this figure we compare the errors in the estimation of $t_c$, 
  $\M$ and $\eta$ for equal-mass black hole
  coalescences  by the initial LIGO interferometers
  at a fixed inspiral SNR of 10. The solid line
  corresponds to the inspiral signal only and the others to the 
  combined inspiral plus ringdown waveforms.
  The dashed line corresponds to the case in which only the five 
  independent inspiral
  parameters (\ref{parameter_inspiral}) are considered for 
   $\epsilon=1.5\%$, while the dot-dashed lines correspond
  to the cases in which we consider all the independent global parameters 
  (\ref{global_parameter_set}) and $\epsilon$ equals to 1.5\% and 0.5\%,
   respectively.}
  \label{f1.IL}
\end{figure}

\begin{figure}
  \begin{center}
  \begin{tabular}{c}
  \includegraphics[height=6cm]{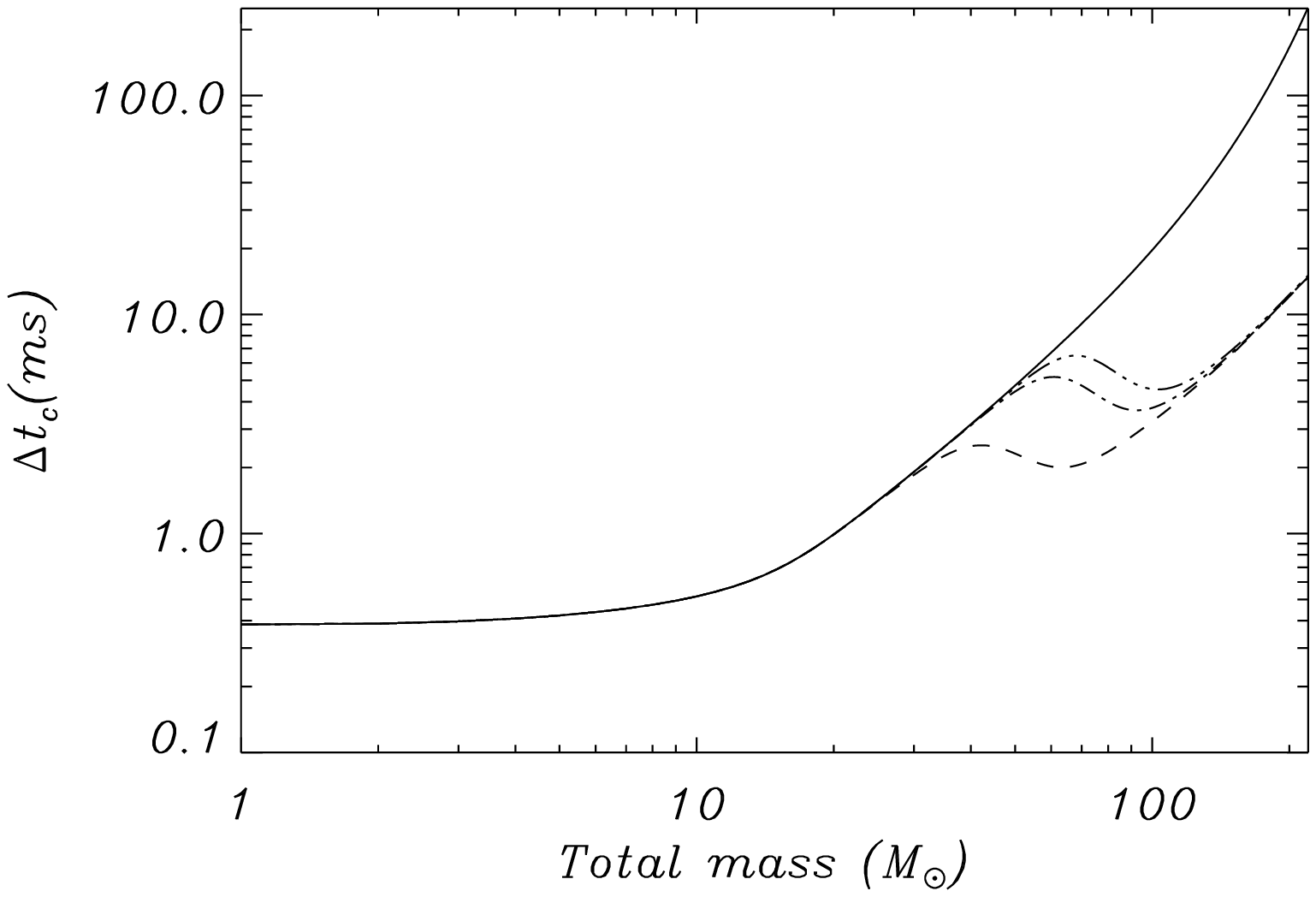}\\
  \includegraphics[height=6cm]{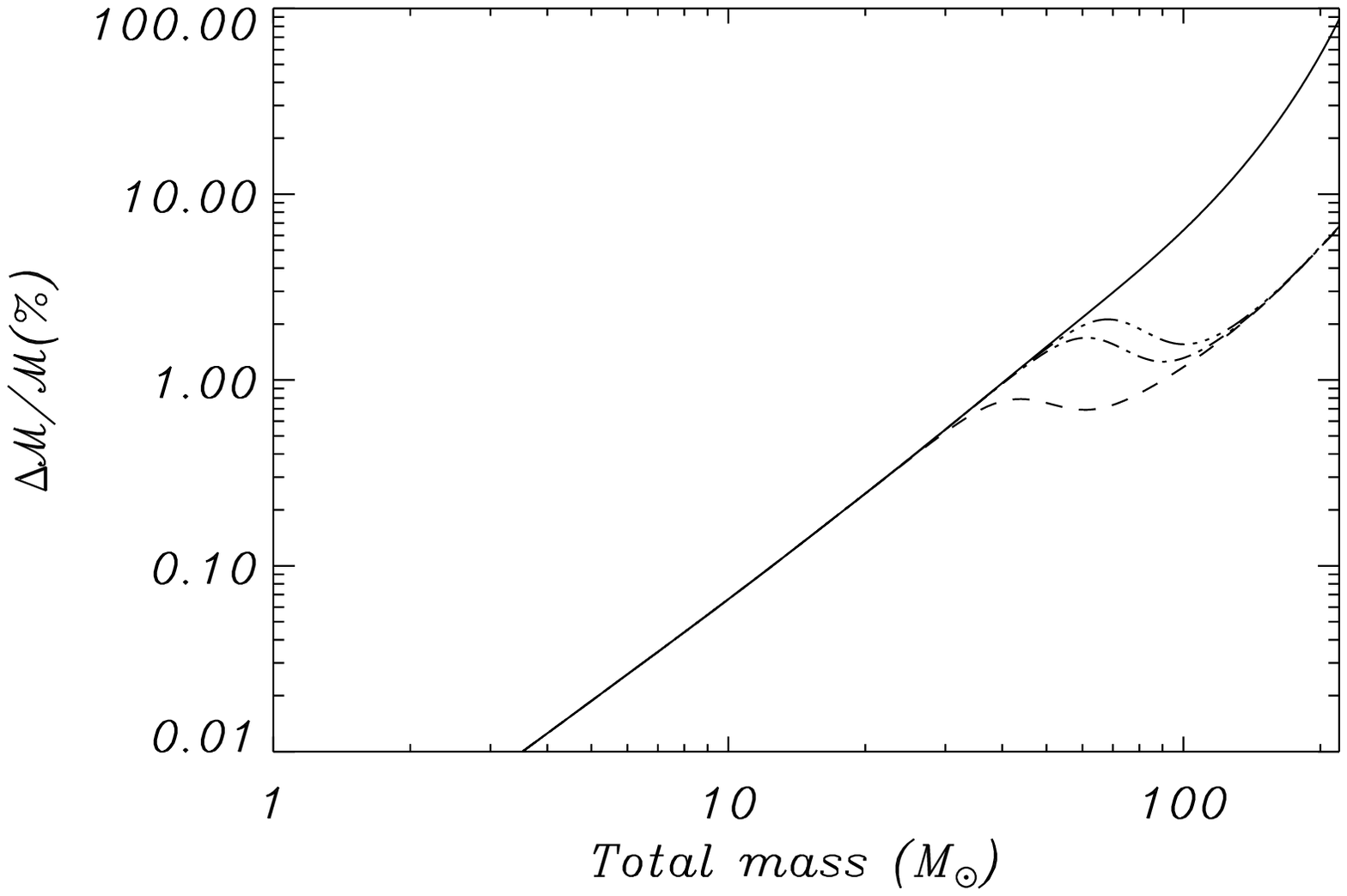} \\ 
  \includegraphics[height=6cm]{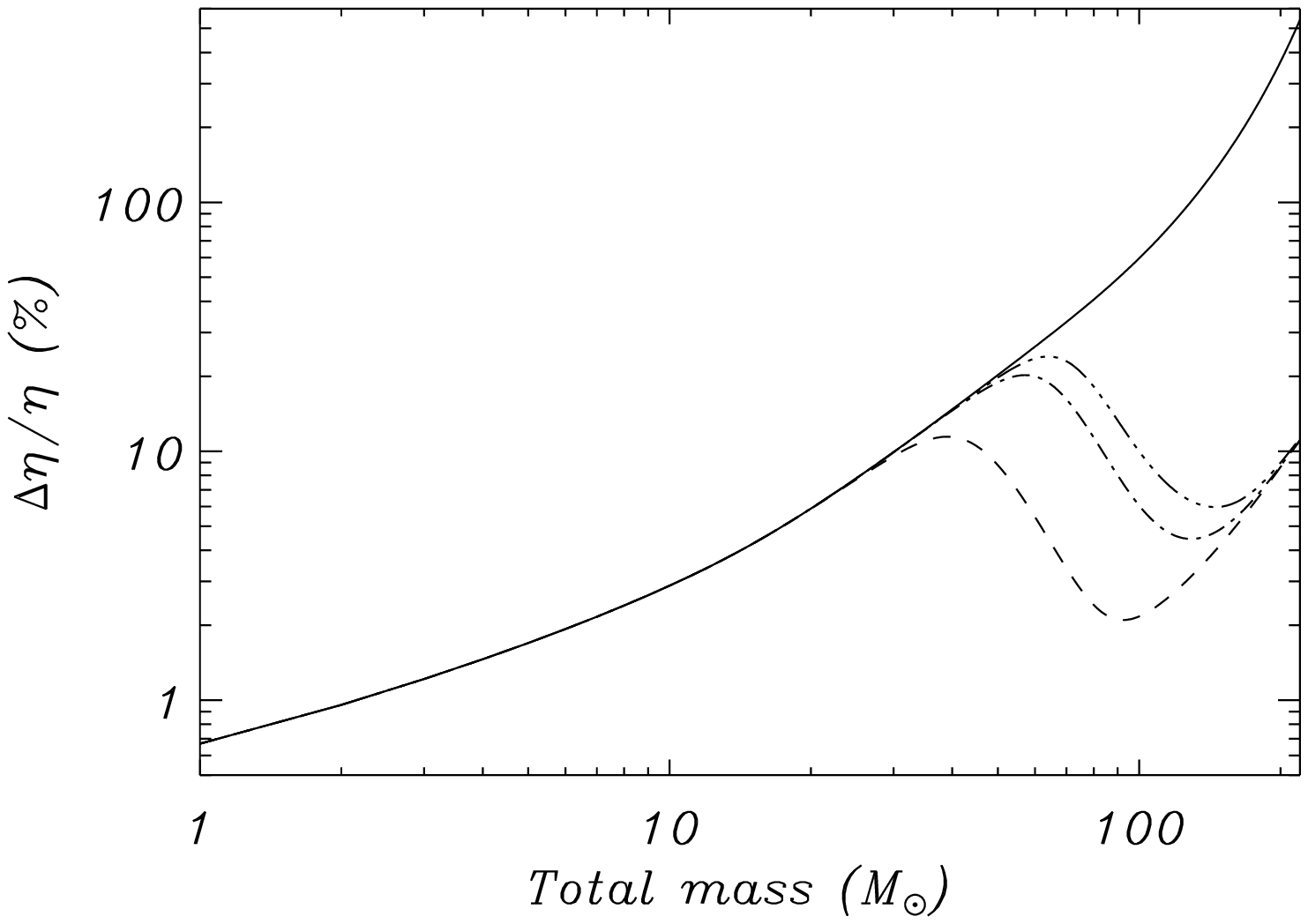}\\
  \end{tabular}  
  \end{center}
  \caption{Same as figure~\ref{f1.IL} for advanced LIGO.}
 \label{f2.AL}
\end{figure}

\begin{figure}
  \begin{center}
  \begin{tabular}{c}
  \includegraphics[height=6cm]{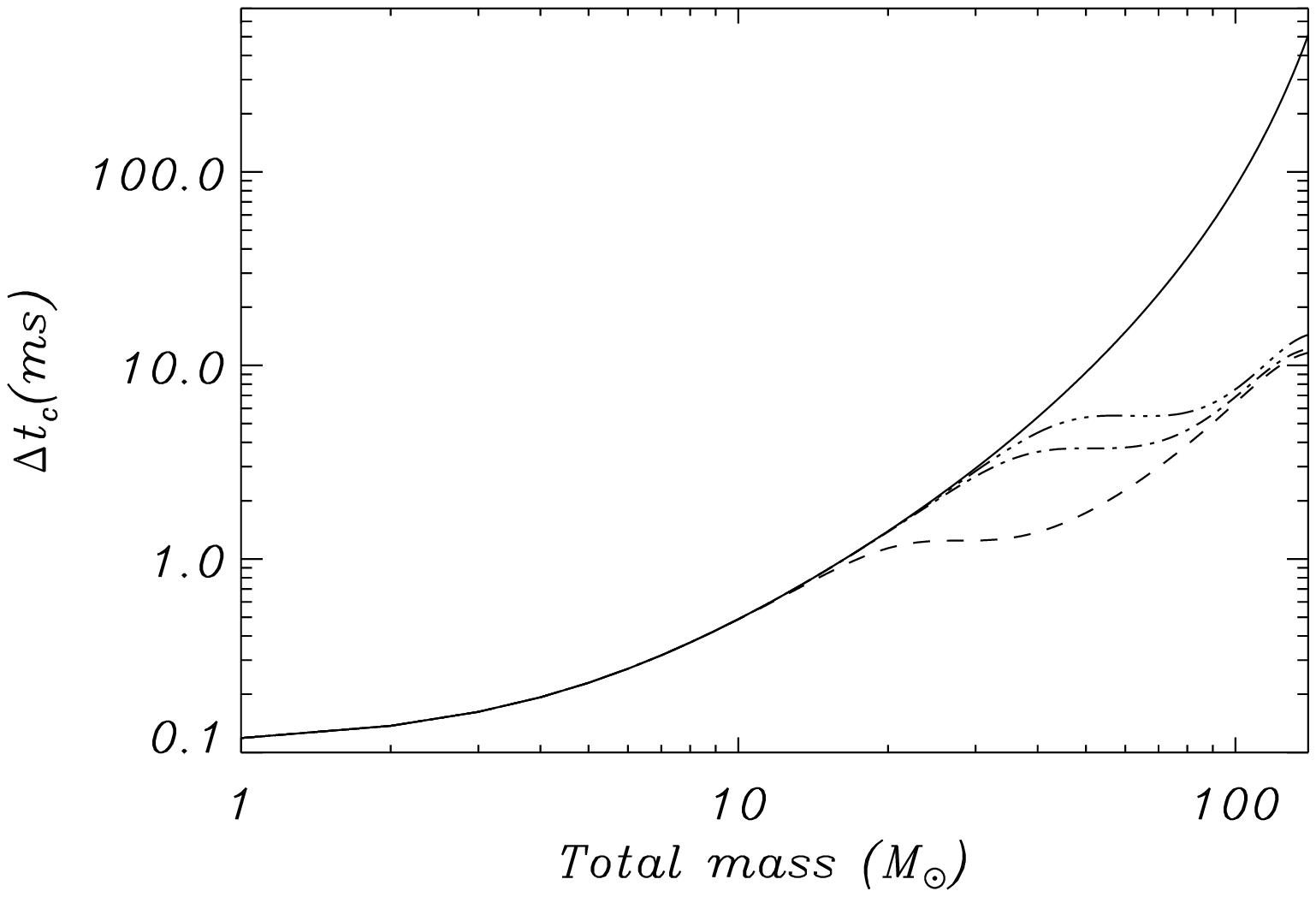}\\
  \includegraphics[height=6cm]{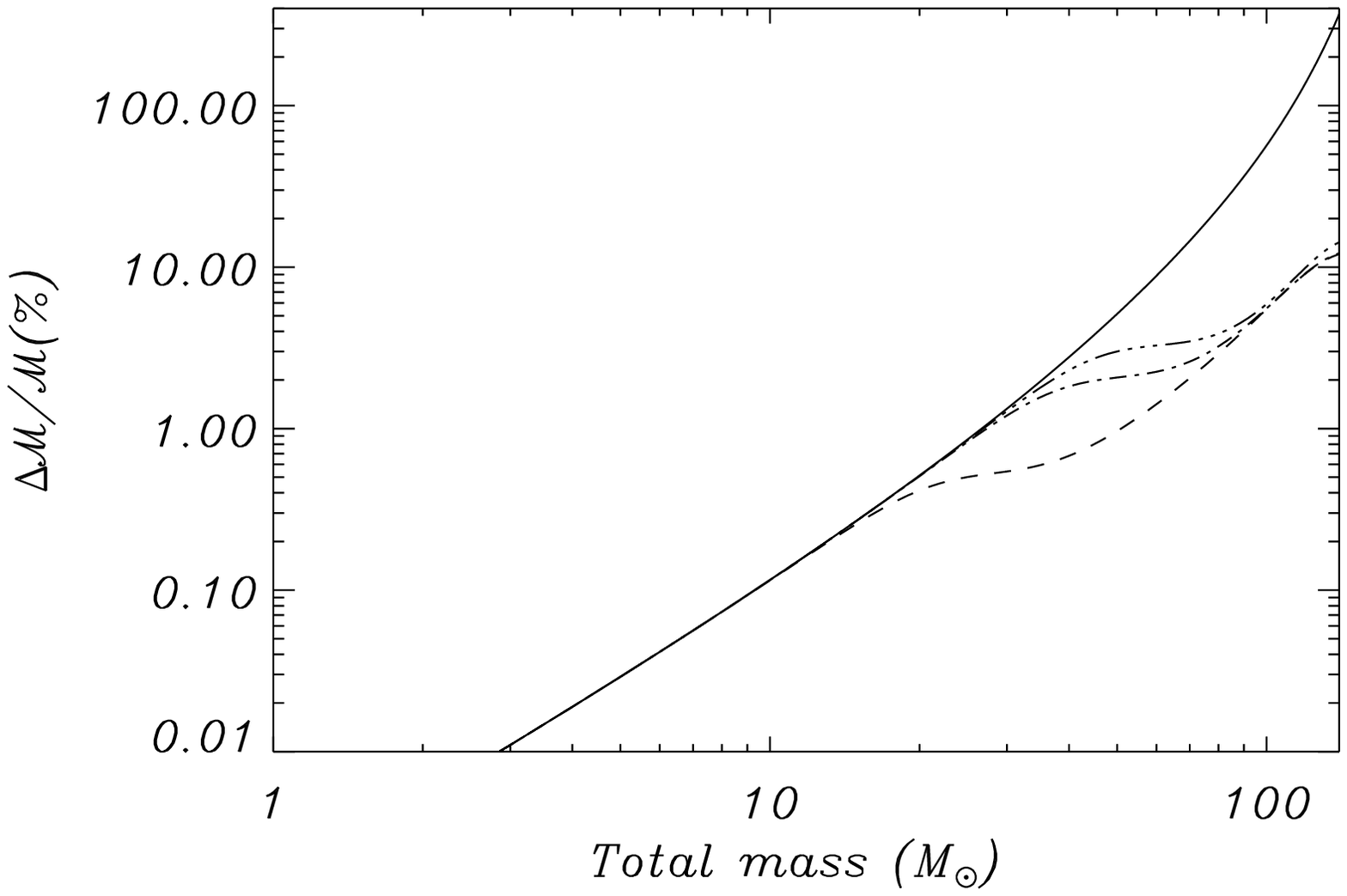} \\ 
  \includegraphics[height=6cm]{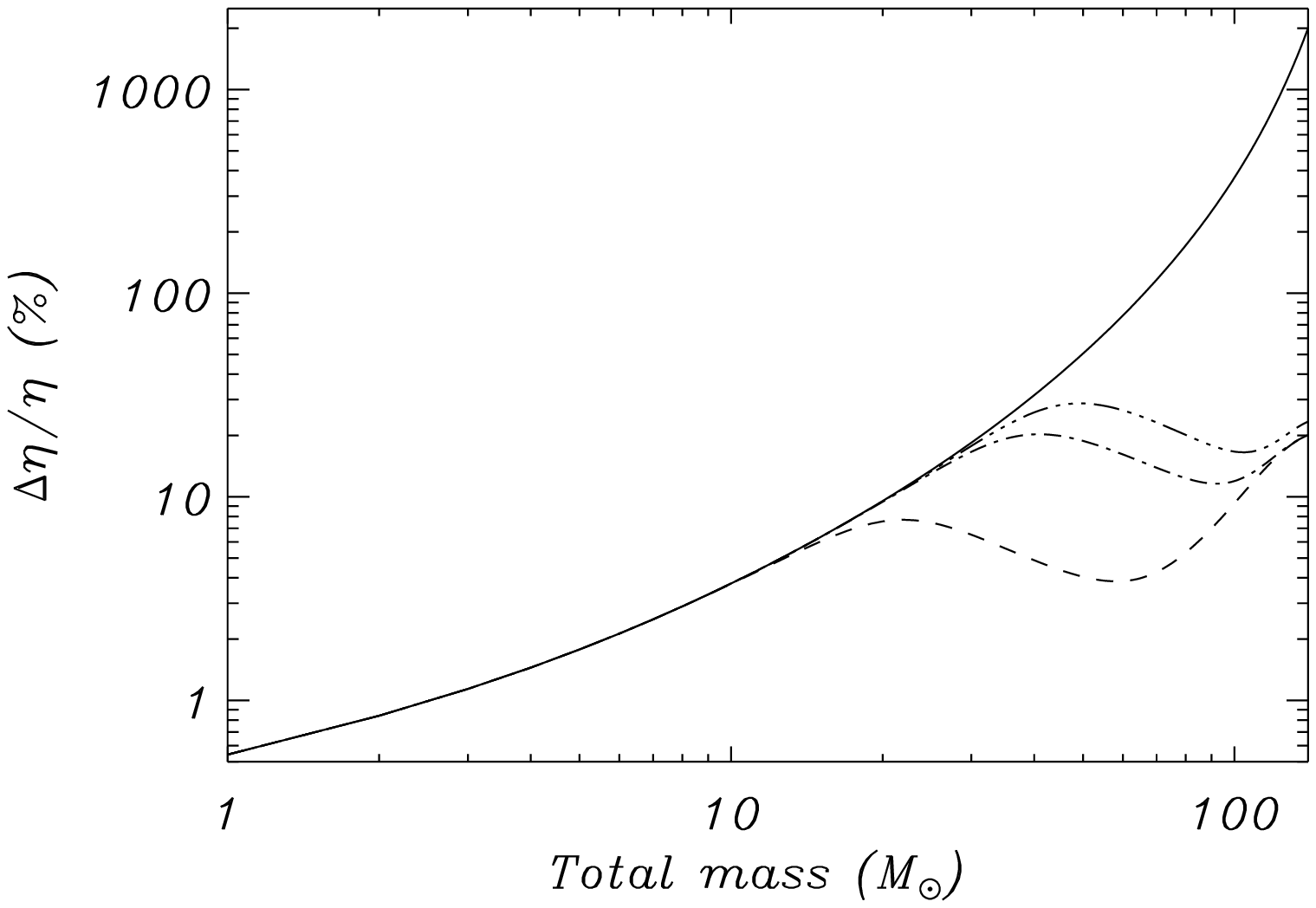}\\
  \end{tabular} 
  \end{center}
  \caption{Same as figure~\ref{f1.IL} for VIRGO}
  \label{f3.V}
\end{figure}

For the equal-mass case the  results are presented in 
figures~\ref{f1.IL}, \ref{f2.AL} and \ref{f3.V}
corresponding to initial LIGO, advanced LIGO and VIRGO, respectively.
The errors  of $t_c$, $\M$ and $\eta$ 
and some of the  associated correlation coefficients
for the inspiral signal alone, for 
different pairs of
masses, can be found in table~\ref{t1}.
 In tables~\ref{t2} and \ref{t3}
one can find the comparison of errors  and correlation coefficients
for the different cases we have analyzed. In particular 
table~\ref{t2} refers to the case (i) in which only the five 
inspiral parameters are
considered and table~\ref{t3} refers to the case (ii) in which we use
the eight global parameters.
In all cases the errors improve, and the improvement is higher for larger 
masses for which the ringdown signal contribution to the SNR increases.
This improvement could be explained by the greater structure and variety
of the global waveform  but also by the variation of some of the correlation 
coefficients, although this is not fully assessed in this paper.
We have just noticed that the correlations coefficients relative to the masses
decrease when the ringdown signal is added, as can be seen in the tables.
We also notice that the improvement is very significant for massive 
systems with very large errors for the inspiral waveform alone. These large
errors  are associated to the small number of useful cycles of the inspiral
signal of these systems~\cite{Arun:2004hn}. Therefore the effect induced in 
parameter estimation due to the inclusion of the ringdown signal could be
understood in terms of additional number of gravitational wave cycles
accumulated.  Although, from the present analysis,
it is not clear which of these considerations is the dominant aspect
to completely understand the variation in parameter estimation observed 
with the global waveform.

\begin{table}
\caption{
Measurements of errors and some of the associated correlation coefficients using the 2nd
post-Newtonian
 binary inspiral waveform at a SNR of 10.
 For each of the three detector noise curves the table
presents $\Delta t_c$ (in msec), $\Delta \phi_c $ (in radians), 
$\Delta{\cal M}/{\cal M}$ and $\Delta{\cal \eta}/{\cal \eta}$ (in percentages).
The cases considered here  correspond to NS-BH and BH-BH binaries of different
masses.
}
\begin{tabular}{ccccccccc}\\
\hline
\hline
$m_1$ & $m_2$ &
$\Delta t_c  $&   $\Delta \phi_c $ &  $\Delta{\cal M}/{\cal M}$ & 
$\Delta{\cal \eta}/{\cal \eta} $&
$C_{t_c \cal M}$& $C_{t_c \eta}$& $C_{\cal M \eta}$\\ 
($M_\odot$) &  ($M_\odot$) & (msec) & (rad) &($\%$) &($\%$) 
\\
\hline
\multicolumn{3}{l}{Initial LIGO} \\\hline
20&1.4&3.3708&7.8897&0.7873&10.6451&0.9294&0.9760&0.9847\\
50&1.4&47.2786&65.2176&6.5073&50.6737&0.9845&0.9953&0.9967\\
20&10&7.6032&15.0441&8.0579&72.3184&0.9573&0.9851&0.9918\\
60&10&253.505&291.931&176.566&1058.45&0.9947&0.9983&0.9990\\
\hline
\multicolumn{3}{l}{Advanced LIGO}\\\hline
25&1.4&1.5726&2.3121&0.0769&2.0304&0.8002&0.9397&0.9319\\
100&1.4&20.5373&11.5863&0.3354&0.9734&0.9426&0.9829&0.9859\\
200&1.4&171.714&59.9235&1.5215&12.2235&0.9836&0.9951&0.9965\\
100&10&24.9199&13.3081&2.5514&26.7760&0.9500&0.9843&0.9886\\
200&10&205.361&69.9572&12.1978&92.3077&0.9855&0.9955&0.9970\\
100&50&58.8914&25.7410&17.1443&133.132&0.9707&0.9901&0.9943\\
175&50&282.241&93.6166&66.6933&434.669&0.9884&0.9962&0.9978\\
\hline
\multicolumn{3}{l}{VIRGO}\\\hline
20&1.4&1.5875&2.8685&0.1339&3.0661&0.8619&0.9499&0.9679\\
100&1.4&89.3534&62.0062&3.1241&24.8609&0.9839&0.9951&0.9966\\
20&10&2.9432&4.3971&1.1712&16.7567&0.9025&0.9630&0.9805\\
50&10&14.9211&14.5300&4.7808&46.7855&0.9564&0.9853&0.9912\\
100&10&128.487&85.3035&28.3745&202.072&0.9874&0.9960&0.9975\\
70&50&198.798&128.504&135.517&804.327&0.9907&0.9968&0.9983\\
90&50&509.639&299.335&338.459&1865.97&0.9948&0.9982&0.9991\\
\hline
\end{tabular} 
\label{t1}
\end{table}


\begin{table}
\small
\caption{
Measurements of errors and associated correlation coefficients using the 2nd
post-Newtonian  binary inspiral waveform at a SNR of 10 together with 
the ringdown waveform, using a set of five parameters
$\{ \ln \mathcal{A}_\ins, f_0 t_c, \phi_c, \ln \mathcal{M}, \ln \eta \}$,
 excluding  $\{\ln a, \gamma_0, t_0 \}$.}
\begin{tabular}{cccccccccccc}\\
\hline
\hline
$\epsilon$ & $m_1$ & $m_2$ &
$\frac{\Delta \cal A_\ins}{\cal A_\ins} $&
$\Delta t_c  $&   $\Delta \phi_c $ &  $\Delta{\cal M}/{\cal M}$ & 
$\Delta{\cal \eta}/{\cal \eta} $&
$C_{t_c \cal M}$& $C_{t_c \eta}$& $C_{\cal M \eta}$&
$\rho_\gl$
\\ 
($\%$) & ($M_\odot$) & ($M_\odot$) & ($\%$) & (msec) & (rad) &($\%$) &($\%$) & 
\\
\hline
\multicolumn{3}{l}{Initial LIGO} \\\hline

1.5  & 20 &  1.4 &9.9990 & 3.3424& 7.8207& 0.7807& 10.5517& 0.9282&0.9756 & 0.9844& 10.0010\\

1.5 &50&1.4&9.9697&14.8477 &19.9568 &2.0873 & 15.5191& 0.8468& 0.9513&0.9680 & 10.0304\\

1.5 &20&10&9.9852&2.5040 & 4.5312& 2.6880& 21.6438& 0.6113& 0.8537& 0.9239&10.0148 \\

1.5 &60&10&9.2208& 9.7683& 1.7464& 10.3674& 17.5337& -0.9594& -0.9369&0.9922 & 10.9880\\

\hline

0.5 &20&1.4&9.9997 & 3.3613& 7.8665&0.7851 & 10.6137& 0.9290& 0.9759& 0.9846&10.0003\\

0.5 &50&1.4& 9.9899& 23.1830& 31.6986& 3.2146&24.6364 &0.9362 & 0.9803& 0.9866& 10.0101\\

0.5 &20&10& 9.9951& 3.7473& 7.1810& 3.9909& 34.4457& 0.8250& 0.9372&0.9660 & 10.0050\\

0.5 &60&10&9.6889 & 10.2048& 2.1410&10.8297 &18.7519 & -0.9490&-0.8894 & 0.9796& 10.3398\\

\hline
\multicolumn{3}{l}{Advanced LIGO}\\\hline

1.5 &25&1.4&9.9994 & 1.5723&2.3116 &0.0769 &2.0300 &0.8002 & 0.9396& 0.9318& 10.0006\\

1.5 &100&1.4&9.5820 &14.3057 & 7.9691&0.2351 & 2.7291& 0.8825& 0.9644& 0.9711&10.4363 \\

1.5 &200&1.4&7.9748 & 27.1408& 8.3128&0.2626 & 1.7010& 0.4037&0.7836 &0.8779 & 12.5470\\

1.5 &100&10& 7.4090& 3.9747&0.9300 & 0.4705&1.4721 &-0.6895 & -0.1562& 0.7144& 13.5008\\

1.5 &200&10& 4.4008& 15.1787& 1.2639& 1.2077& 2.1026&-0.9234 &-0.8621 & 0.9726& 23.1557\\

1.5 &100&50& 3.9049& 6.6557&0.8229 & 2.2776& 3.8235& -0.8769& -0.8672& 0.9951& 28.3348\\

1.5 &175&50& 4.7381& 16.8821& 1.1369& 5.3223& 8.8739&-0.9333 & -0.9325&0.9997 & 47.1507\\

\hline

0.5 &25&1.4&9.9998 & 1.5725& 2.3119&0.0769 & 2.0302&0.8002 &0.9397 & 0.9319& 10.0002\\

0.5 &100&1.4&9.8546 & 17.6073&9.8910 & 0.2882&3.3904 &0.9221 &0.9766 &0.9809 &10.1475 \\

0.5 &200&1.4& 9.1638& 42.2225& 14.0443& 0.3882&2.8678 & 0.7397& 0.9155&0.9452 &10.9152 \\

0.5 &100&10& 8.8592& 4.2986&1.2823 &0.4993 &2.3083 &-0.4740 & 0.2010&0.6867 & 11.2882\\

0.5 &200&10& 6.4175&15.3235 &1.4330 &1.2144 & 2.2822&-0.8994 & -0.7383&0.9295 &15.6651 \\

0.5 &100&50& 5.6355& 6.6795&0.8354 &2.2862 & 3.8925&-0.8744 &-0.8461 &0.9858 & 18.2835\\

0.5 &175&50& 5.2676&17.0019 & 1.1474& 5.3641& 8.9507& -0.9338&-0.9315 & 0.9991& 28.4206\\

\hline

\multicolumn{3}{l}{VIRGO}\\\hline

1.5 &20&1.4&9.9884 & 1.5739&2.8419 &0.1328 &3.0372 &0.8595 & 0.9490& 0.9673&10.0116 \\

1.5 &100&1.4& 9.8123& 23.6168& 15.7433&0.8518 &6.3184 & 0.7771&0.9281 &0.9533 & 10.1916\\

1.5 &20&10& 9.8860& 1.3595& 1.8045&0.5264 &6.7294 & 0.5307&0.8124 &0.8998 & 10.1154\\

1.5 &50&10&9.5906 &2.5435 &1.4330 &0.8642 &4.2425 &-0.4119 &0.2591 &0.7145 & 10.4273\\

1.5 &100&10& 8.7416& 8.8760&1.5912 & 2.6407&4.9586 &-0.8994 &-0.7351 &0.9318 & 11.4628\\

1.5 &70&50& 8.0221& 10.4639& 1.1242& 9.8663&16.4643 &-0.9298 &-0.9263 &0.9987 &15.1284\\

1.5 &90&50&8.8177 &12.3529 & 1.0874&11.8368 &19.7325 &-0.9136 &-0.9119 & 0.9994& 18.3197\\

\hline

0.5 &20&1.4& 9.9961&1.5830 & 2.8595& 0.1335& 3.0564&0.8611 &0.9496 &0.9677 &10.0039 \\

0.5 &100&1.4&9.9362 & 37.4447&25.6283 & 1.3239& 10.2789&0.9094 &0.9719 &0.9808 &10.0643 \\

0.5 &20&10&9.9616 &1.8724 & 2.6772& 0.7371&10.1277 &0.7565 &0.9058 & 0.9501& 10.0386\\

0.5 &50&10&9.8577 & 3.1084& 2.2677& 1.0361& 7.0748&0.0358 &0.5871 &0.7987 & 10.1444\\

0.5 &100&10& 9.5177&9.1524 & 2.1289& 2.6903&5.9890 &-0.8220 &-0.4653 &0.8587 & 10.5102\\

0.5 &70&50&8.7160 &10.3890 & 1.2208& 9.7697&16.3850 &-0.9375 &-0.9251 &0.9955 & 11.9564\\

0.5 &90&50&9.0164 &14.0215 &1.1789 &13.7140 &22.8751 &-0.9326 & -0.9288& 0.9987& 13.3618\\

\hline
\end{tabular} 
\label{t2}
\end{table}

\begin{table}
\small
\caption{
Measurements of errors and associated correlation coefficients using the 
2nd
post-Newtonian  binary inspiral waveform at a SNR of 10 together with 
the ringdown waveform, using a set of eight parameters
$\{ \ln \mathcal{A}_\ins, f_0 t_c, \phi_c, \ln \mathcal{M}, \ln \eta, 
\ln a, \gamma_0, t_0 \}$. }
\begin{tabular}{cccccccccccc}\\
\hline
\hline
$\epsilon$&
$m_1$&
$m_2$&
$\frac{\Delta \cal A_\ins}{\cal A_\ins} $&
$\Delta t_c  $&  
$\Delta \phi_c $ &  
$\Delta{\cal M}/{\cal M} $ & 
$\Delta{\cal \eta}/{\cal \eta} $&
$C_{t_c \cal M}$&
$C_{t_c \eta}$&
$C_{\cal M \eta}$&
$\rho_\gl$
\\ 
($\%$) & ($M_\odot$)  & ($M_\odot$) & ($\%$) & (msec) & (rad) &($\%$) &($\%$) & & 
\\
\hline
\multicolumn{3}{l}{Initial LIGO} \\\hline
1.5  & 20 & 1.4 &9.9991 & 3.3699& 7.8874& 0.7871& 10.6420&0.9293 &0.9760 &0.9847 & 10.0010\\

1.5  & 50 & 1.4 &9.9726 & 41.1566&56.7201 & 5.6682&44.0711 &0.9796 & 0.9938&0.9957 & 10.0304\\

1.5 &20&10 & 9.9867& 6.5956&13.0066 &6.9918 & 62.5063&0.9433 &0.9801 &0.9890 & 10.0148\\

1.5 &60&10 &9.3295 &11.0001 & 4.9916& 10.5193&22.7730 & -0.7781& -0.3566&0.8522 & 10.9880\\

\hline
0.5 &20&1.4 &9.9997 & 3.3705&7.8889 & 0.7872&10.6441 & 0.9294&0.9760 & 0.9847& 10.0003\\

0.5 &50&1.4 &9.9907 & 44.9413& 61.9743&6.1869 & 48.1536&0.9829 &0.9948 & 0.9964& 10.0101\\

0.5 &20&10 &9.9955 & 7.2160& 14.2620&7.6482 & 68.5521&0.9526 &0.9834 & 0.9908& 10.0050\\

0.5 &60&10 &9.7394 & 12.6647& 8.1550& 11.5177& 32.7242&-0.5226 & 0.0717& 0.8049& 10.3398\\

\hline
\multicolumn{3}{l}{Advanced LIGO}\\\hline
1.5 &25&1.4 & 9.9995&1.5726 &2.3120 &0.0769 & 2.0304&0.8002 &0.9397 & 0.9319& 10.0006\\

1.5 &100&1.4 & 9.6291&20.2342 & 11.4112& 0.3305& 3.9132& 0.9409&0.9824 & 0.9855& 10.4363\\

1.5 &200&1.4 &8.1761 &100.5296 &34.8762 & 0.8944& 7.1141&0.9524 &0.9855 & 0.9898& 12.5470\\

1.5 &100&10 & 7.6653& 7.6876&3.6663 &0.8168 & 7.2850&0.4943 &0.8220 &0.8837 & 13.5008\\

1.5 &200&10 & 4.7707& 16.9594&2.8159 & 1.2761& 3.8796& -0.6381&-0.0433 & 0.7703& 23.1557\\

1.5 &100&50 &4.1663 & 6.7824&0.9771 & 2.2953& 4.5952&-0.8306 & -0.6018& 0.8899& 28.3348\\

1.5 &175&50 & 4.8136& 16.9720& 1.2004&5.3227 &8.9546 &-0.9283 &-0.9057 &0.9909 & 47.1507\\

\hline
0.5 &25&1.4 &9.9998 & 1.5726& 2.3121& 0.0769& 2.0304& 0.8002& 0.9397& 0.9319&10.0002 \\

0.5 &100&1.4 & 9.8716& 20.4346& 11.5270& 0.3338&3.9530 &0.9421 & 0.9827& 0.9858& 10.1475\\

0.5 &200&1.4 &9.2305 &132.9684 & 46.3067& 1.1798&9.4457 &0.9727 &0.9917 &0.9942 & 10.9152\\

0.5 &100&10 & 8.9967&11.4032 &5.8388 & 1.1846& 11.6961&0.7646 &0.9230 &0.9461 &11.2882 \\

0.5 &200&10 &6.8066 &19.8847 &4.4807 & 1.4050& 5.9807&-0.2786 & 0.3719& 0.7714& 15.6651\\

0.5 &100&50 & 5.9767& 7.0260& 1.2240& 2.3409&5.8647 & -0.7460& -0.3025&0.7993 & 18.2835\\

0.5 &175&50 &5.4400 & 17.2493&1.3221 &5.3654 &9.1950 & -0.9194& -0.8559& 0.9743& 28.4206\\

\hline
\multicolumn{3}{l}{VIRGO}\\\hline
1.5 &20&1.4 &9.9894 &1.5871 & 2.8676&0.1339 &3.0652 &0.8619 &0.9498 & 0.9679& 10.0116\\

1.5 &100&1.4 & 9.8253&71.2674 & 49.3691&2.4945 & 19.7938& 0.9747&0.9923 &0.9946 &10.1916 \\

1.5 &20&10 & 9.8943& 2.7362& 4.0693&1.0874 &15.4957 &0.8870 & 0.9570& 0.9774& 10.1154\\

1.5 &50&10 &9.6406 & 6.5355& 6.0863& 2.1034& 19.5048& 0.7740&0.9210 & 0.9537& 10.4273\\

1.5 &100&10 &8.9350 &11.7992 & 5.2447&3.0553 & 12.5793& -0.2535&0.3875 & 0.7797& 11.4628\\

1.5 &70&50 &8.0194 & 10.1883& 1.6360&9.0353 &15.5288 & -0.9005& -0.7771& 0.9528& 15.1284\\

1.5 &90&50 & 8.8495& 12.9812& 1.4506& 11.9826&19.7503 &-0.9062 &-0.8606 & 0.9839& 18.3197\\

\hline

0.5 &20&1.4 & 9.9964&1.5874 &2.8682 & 0.1339&3.0658 &0.8619 & 0.9499&0.9679 &10.0039 \\

0.5 &100&1.4 &9.9386 &81.8448 & 56.7624& 2.8626& 22.7583&0.9808 & 0.9942& 0.9959&10.0643 \\

0.5 &20&10 &9.9641 & 2.8685&4.2790 & 1.1410&16.3023 & 0.8973& 0.9610&0.9795 &10.0386 \\

0.5 &50&10 &9.8721 & 9.4181&9.0312 &3.0223 &29.0276 & 0.8908&0.9627 & 0.9778&10.1444 \\

0.5 &100&10 & 9.6003&15.8131 &8.7251 &3.8272 & 20.7374& 0.2455& 0.7029& 0.8553& 10.5102\\

0.5 &70&50 &8.8855 &11.2685 & 2.2537& 9.7760& 18.2478& -0.8595& -0.5971&0.9003 & 11.9564\\

0.5 &90&50 &9.0797 &15.2788 & 1.9791&13.9267 & 22.9881& -0.9111&-0.8146 &0.9655 & 13.3618\\

\hline
\end{tabular} 
\label{t3}
\end{table}


The numerical results for the inspiral and the ringdown waveforms separately
have been verified by comparing with those existing in the literature
(for different masses and noise curves). Moreover, for the ringdown case alone
 we have also found a good agreement 
 with  an analytical approximation
 as described in~\ref{B}.

\section{Conclusions}
\label{sec:conc}

We have carried out a study to understand the implications of adding the
ringdown to the inspiral signal on parameter estimation of non-spinning binaries
using the covariance matrix. We have compared the results using three different noise
curves corresponding to initial LIGO, advanced LIGO and VIRGO.  

The result of our study is that the parameter estimation of $t_c$, $\phi_c$, 
${\cal M}$ and $\eta$ improves significantly,
as expected, by employing the extra information that comes from the ringdown,
for those systems with a total mass such that both the inspiral and the
 ringdown signal could be detectable by the detectors. Naturally the improvement
 is larger in the case of considering a smaller number of parameters, but in
 both cases, the five parameter case and the eight parameter case, 
 the improvement is significant.
  The study is performed at a fixed inspiral SNR of 10, therefore the error in
 ${\Delta \cal A_\ins}/{\cal A_\ins} $ would be of
  $10\%$ for the inspiral signal alone. This is also improved by adding the
  ringdown.

In this work we have made a number of simplifying assumptions, ignoring the
merger phase, considering only the 2nd post-Newtonian inspiral phase
formula instead of the 3.5 that is already known, using only a single mode
for the ringdown signal,
 and ignoring angular dependencies (because of the angle
averages we use). For this reason  the results obtained here should be
considered just as an indication of which could be the real effect
 in the parameter estimation by combining the inspiral with the 
 ringdown signal. The preliminary results obtained here seem to be
 very encouraging. Therefore  it
 would be interesting to extend the analysis to a more
 realistic case  and also for different data analysis techniques 
 (different from  matched filtering).
 Although  we have focused on ground-based detectors a similar study could be
 performed for LISA.

\section*{Acknowledgments}
The authors gratefully acknowledge the support of 
the Spanish Ministerio de Educaci\'on y Ciencia research project 
FPA-2004-03666.

\appendix
\section{The waveform derivatives}
\label{derivatives}

In order to calculate the Fisher matrix with respect to the 
$\{ \ln \mathcal{A}_\ins$, $f_0 t_c$, $\phi_c$, $\ln \mathcal{M}$,
$\ln \eta$, $\ln a$, $\gamma_0$, $t_0 \}$
basis, we need to compute first the waveform derivatives. For the inspiral 
waveform $\tilde{h}_\ins(f)$ these are the following
\begin{eqnarray}
\label{derivative_amplitude_inspiral}
{\partial \tilde{h}_\ins \over \partial \ln \mathcal{A}_\ins}&=& \tilde{h}_\ins\\
\label{derivative_t0_inspiral}
{\partial \tilde{h}_\ins \over \partial f_0 t_c}&=&
i \; 2 \pi (f/f_0) \tilde{h}_\ins\\
\label{derivative_phase_c_inspiral}
{\partial \tilde{h}_\ins \over \partial \phi_c}&=&
- i \; \tilde{h}_\ins\\
\label{derivative_chirpmass_inspiral}
{\partial \tilde{h}_\ins \over \partial \ln \mathcal{M}}&=& i \; 
\frac{1}{128}
\sum_{k=0}^4 A_k ~ (k-5) u^{k-5} \; \tilde{h}_\ins\\
\label{derivative_eta_inspiral}
{\partial \tilde{h}_\ins \over \partial \ln \eta}&=&
i \; \frac{3}{128} \sum_{k=0}^4 B_k ~ u^{k-5} \;\tilde{h}_\ins \, ,
\end{eqnarray}
%
%
where the parameter  $A_k$ are given by equations
(\ref{alpha_parameters})-(\ref{alpha_parameters2}) and  $B_k$ are
\begin{eqnarray}
B_0 &=& \frac{\partial A_0}{\partial \ln \eta}=0\\
B_1 &=& \frac{\partial A_1}{\partial \ln \eta}=0\\
B_2 &=& \frac{\partial A_2}{\partial \ln \eta}=(-\frac{743}{378}+\frac{11}{3}
\eta)\eta^{-2/5}\\
B_3 &=& \frac{\partial A_3}{\partial \ln \eta}=\frac{48}{5} \pi \eta^{-3/5}\\
B_4 &=& \frac{\partial A_4}{\partial \ln \eta}=(-\frac{3058673}{127008}+
\frac{5429}{504}\eta
+\frac{617}{12}\eta^2)\eta^{-4/5} \, .
\end{eqnarray}
The inspiral waveform has no dependency on $\ln a$, $ \gamma_0$ and $t_0$.
Therefore the remaining derivatives vanish
\be
{\partial \tilde{h}_\ins \over \partial \ln a}=
{\partial \tilde{h}_\ins \over \partial \gamma_0} =
{\partial \tilde{h}_\ins \over \partial t_0}= 0 \, .
\ee

The derivatives of the ringdown waveform $\tilde{h}_\rd (f)$
can be computed by taking into account the  implicit dependencies of 
$\mathcal{A}_\rd(\mathcal{A}_\ins,\mathcal{M},\eta, a)$,
$f_\qnr(\mathcal{M},\eta,a)$, and $Q(a)$. We get
\begin{eqnarray}
\frac{\partial \tilde{h}_\rd}{\partial \ln \mathcal{A}_\ins} &=&
\tilde{h}_\rd\\
\frac{\partial \tilde{h}_\rd}{\partial \ln \mathcal{M}} 
&=& \frac{1}{6} ~\tilde{h}_\rd-
\frac{\partial \tilde{h}_\rd}{\partial \ln f_\qnr}\\
\frac{\partial \tilde{h}_\rd}{\partial \ln \eta} &=&
\frac{2}{5} ~\tilde{h}_\rd+ \frac{3}{5}
\frac{\partial \tilde{h}_\rd}{\partial \ln f_\qnr}\\
\frac{\partial \tilde{h}_\rd}{\partial \ln a} &=&
\frac{9(-100+21(1-a)^{3/10}a)}{40(-100+63(1-a)^{3/10})(-1+a)} ~\tilde{h}_\rd
+{}\nonumber\\
& & {}+\frac{189 a}{10(-63+100(1-a)^{7/10}+63a)}
\frac{\partial \tilde{h}_\rd}{\partial \ln f_\qnr}+{}\nonumber\\
& &{} +\frac{9 a}{20(1-a)}
\frac{\partial \tilde{h}_\rd}{\partial \ln Q}
\end{eqnarray}
where
\bea
\frac{\partial \tilde{h}_\rd}{\partial \ln f_\qnr} &=& 
\frac{\mathcal{A}_\rd e^{2 i f \pi  t_0} f_{\qnr}}{2 \pi} \\
&  \times &
\left(\frac{e^{i
\gamma_0 } (2 i Q+1) Q}{(f_{\qnr} (i-2 Q)+2 f
   Q)^2}+\frac{e^{-i \gamma_0 } (1-2 i Q) Q}{(2 f Q+f_{\qnr} (2 Q+i))^2}\right)
 \nonumber
 \eea
 \bea
\frac{\partial \tilde{h}_\rd}{\partial \ln Q}&=& 
\frac{\mathcal{A}_\rd e^{2 i f \pi  t_0} f_{\qnr} }{2 \pi  Q} \\
&  \times &
\left(\frac{e^{i \gamma_0
}}{\left(f_{\qnr} \left(2 i+\frac{1}{Q}\right)-2 i
   f\right)^2}+\frac{e^{-i \gamma_0} }
{ \left( \frac{f_{\qnr}}{Q}-2 i (f+f_{\qnr}) \right)^2}\right) \nonumber
\eea
The remaining derivatives are
\bea
\frac{\partial \tilde{h}_\rd}{\partial \gamma_0} &=& 
\frac{\mathcal{A}_\rd e^{2 i f \pi  t_0} }{2 \pi } \\
& & \times
\left(\frac{i e^{i \gamma_0
}}{\frac{f_{\qnr}}{Q}-2 i
   (f-f_{\qnr})}-\frac{i e^{-i \gamma_0 }}
   {\frac{f_{\qnr}}{Q}-2 i (f+f_{\qnr})}\right)
   \nonumber\\
\frac{\partial \tilde{h}_\rd}{\partial t_0} &=& 2 i f \pi \tilde{h}_\rd
\eea
\be
{\partial \tilde{h}_\rd \over \partial f_0 t_c}=
{\partial \tilde{h}_\rd \over \partial \phi_c}= 0
\ee

\section{Analytical analysis of the Fisher matrix for the ringdown waveform} 
\label{B}

In what follows we are interested in finding an analytical approximation
to the  Fisher matrix for the ringdown waveform in order to compare and verify
the numerical results obtained with a Fortran code. For this comparison let
 us focus with the simpler case with  five parameters,
in which we are interested in computing the Fisher matrix with respect to
the basis $( \ln \mathcal{A}_\ins, f_0 t_c, \phi_c, 
\ln \mathcal{M}, \ln \eta )$, thus  assuming that there are no uncertainties
in the parameters  $(\ln a, \gamma_0, t_0)$. Of course, the 
ringdown signal does not depend on  $f_0 t_c$  and $\phi_c$, 
therefore the problem is reduced  to  three parameters, although 
the signal would depend only on two
independent  ones, e.g.,  $(\ln \mathcal{A}_\rd, \ln f_\qnr)$.

The way  we proceed  is to compute first the Fisher 
matrix of the ringdown signal
with respect to  $(\ln \mathcal{A}_\rd, \ln f_\qnr)$.
Assuming constant noise over the bandwidth of the 
signal\footnote{The approximation that the noise is constant 
over the bandwidth  of the signal is a good approximation for 
all the detectors considered here when $a\ge 0.9$ corresponding to 
$\Delta f/ f_\qnr \le 0.5$ as explained 
in~\cite{Finn:1992wt}. In this paper we consider only the case in which 
$a=0.98$.}, and taking the
$\delta$-function approximation as in~\cite{Finn:1992wt}, the elements
$\Gamma_{\ln \mathcal{A} \, \ln \mathcal{A}}$ and 
$\Gamma_{\ln f_\qnr\, \ln f_\qnr}$ can be computed 
analytically using Mathematica. For $\gamma_0=0$ we 
get\footnote{If instead of using  $\gamma_0=0$ we take $\gamma_0=\pi /2$ then
$\Gamma_{\ln \mathcal{A}_\rd \, \ln \mathcal{A}_\rd}$ becomes
 $ 4 \mathcal{A}_\rd^2 \, Q^3 / [ \pi (1+ 4 Q^2)f_\qnr S_n (f_\qnr) ]$ 
 equivalent to Finn's result~\cite{Finn:1992wt}.}:
\be
\label{snr2_finn}
\Gamma_{\ln \mathcal{A}_\rd \, \ln \mathcal{A}_\rd}=
\frac{2 \mathcal{A}_\rd^2 \, Q (1+ 2Q^2)}
{ \pi (1+ 4 Q^2)f_\qnr S_n (f_\qnr) } \, ,
\ee
and
\be
\Gamma_{\ln f_\qnr\, \ln f_\qnr}=
\frac{ \mathcal{A}_\rd^2 \, Q (1+ 4Q^2+ 8Q^8)}
     { \pi (1+ 4 Q^2) f_\qnr S_n (f_\qnr) } \, .
\ee
For the cross term  $\Gamma_{\ln \mathcal{A} \, \ln f_\qnr}$, Finn's
approximation \cite{Finn:1992wt} can no longer be employed, because 
$\partial \tilde{h}_\rd / \partial \ln f_\qnr$  is not a symmetric function
around $f_\qnr$. In order to compute this term we will consider the 
following properties we have derived.


The reader should notice that for any set of parameters
 $(\ln \mathcal{A}, \{\lambda^i\})$ and any waveform of the form
\be
\tilde h ( \mathcal{A}, \{\lambda^i\}, f) = 
  \mathcal{A} \,\tilde H(\{\lambda^i\}, f)  \, ,
\ee
the elements of the Fisher matrix  satisfy the relations
\be
\Gamma_{\ln \mathcal{A} \,\lambda^i} 
=\frac{1}{2}{\partial \, \Gamma_{\ln \mathcal{A} \, \ln \mathcal{A}}
\over \partial \lambda^i} \, ,
\ee
\be
\label{ij}
\Gamma_{i j}= \partial_j \Gamma_{\ln \mathcal{A} \, \lambda^i}- (h |
\partial_{i j} h ) \, .
\ee
Then using the standard definition of SNR given by Eq.~(\ref{snr_2})
we have
\be
\label{11}
\Gamma_{\ln \mathcal{A} \, \ln \mathcal{A}} = \rho^2 \, ,
\ee
and consequently
\be
\label{1i}
\Gamma_{\ln \mathcal{A} \,\lambda^i} =\frac{1}{2}\partial_i \rho^2 \, .
\ee
These relations hold true for both the inspiral and the ringdown signals
when considering $\mathcal{A}$ to be the amplitude of the signal.
In case of the ringdown signal, using equations (\ref{snr2_finn}), (\ref{11})
 and (\ref{1i}) we get
\be
\rho_\rd^2=
\frac{2 \mathcal{A}_\rd^2 \, Q (1+ 2Q^2)}
    { \pi (1+ 4 Q^2) f_\qnr S_n (f_\qnr)} \, ,
\ee
\be 
\Gamma_{\ln \mathcal{A}_\rd \ln f_\qnr}=
 -\frac{1}{2} \, \rho_\rd^2  \, \left( 1+ S \right) \, ,
 \ee
where 
\be
S=  \frac{1}{S_n(f_\qnr)}
\frac{d \, S_n(f_\qnr)}{d \, \ln f_\qnr} \, .
\ee
Note that the $\delta$-function approximation in this case is equivalent to consider $S=0$,
but this term is not negligible. For example if we consider initial LIGO
and a total mass of 10, 20 or 100 $M_\odot$, the corresponding $S$ value
would be 1.989, 1.959 and 1.264 respectively. 

The Fisher matrix with   respect to the basis 
  $( \ln \mathcal{A}_\ins, \ln \mathcal{M}, \ln \eta )$ (which 
naturally will be degenerate) can be easily be computed by 
taking into account the amplitude relation given by 
Eq.~(\ref{amplitudes_relation}) and considering 
\be
\left( \begin{array}{c}
{{\normalsize\partial \tilde h_\rd}} /{ \partial \ln \mathcal{A}_\ins }\\
{\normalsize\partial \tilde h_\rd}/ { \partial \ln \mathcal{M} }\\
{\normalsize\partial \tilde h_\rd}/ { \partial \ln \eta } \\
\end{array}\right)= 
\left( \begin{array}{cc}
1 & 0 \\
1/6 & -1 \\
2/5 & 3/5 \\
\end{array}
\right) \,
\left( \begin{array}{c}
{\partial \tilde h_\rd } /{ \partial \ln \mathcal{A}_\rd }\\
{\partial \tilde h_\rd }/ { \partial \ln f_\qnr}\\
\end{array}\right) \, .
\ee
If we define the constant matrix 
\be
\mathcal{C} \equiv
\left( \begin{array}{cc}
1 & 0 \\
1/6 & -1 \\
2/5 & 3/5 \\
\end{array}
\right) \, ,
\ee
let $\Gamma$ be the Fisher matrix with respect to  $(\ln \mathcal{A}_\rd, \ln f_\qnr)$
and $\hat\Gamma$ the  Fisher matrix with respect to 
  $( \ln \mathcal{A}_\ins, \ln \mathcal{M}, \ln \eta )$, in this particular
  case,    $\Gamma$ and $\hat\Gamma$ are related in
  the following way
  \be
  \hat\Gamma = \mathcal{C} \, \Gamma \, \mathcal{C}^T \, ,
  \ee
where the superscript $T$ indicates transposed matrix.  The matrix
 $\hat\Gamma$ has the elements:
 
 \be 
\hat\Gamma_{\ln \mathcal{A}_\ins \,\ln \mathcal{A}_\ins}=
  \rho_\rd^2 \, ,
 \ee
 
 \be 
\hat\Gamma_{\ln \mathcal{A}_\ins \ln \mathcal{M}}=
 \rho_\rd^2  \, \frac{4+3S}{6}  \, ,
 \ee
  
 \be 
\hat\Gamma_{\ln \mathcal{A}_\ins \ln \eta}=
 \rho_\rd^2  \, \frac{1-3S}{10}  \, ,
 \ee

 \be 
\hat\Gamma_{\ln \mathcal{M} \ln \mathcal{M}}=
 \rho_\rd^2  \, \frac{2 \left(72 Q^2+6 S+43\right) Q^2+6 S+25}{72 Q^2+36}
  \, ,
 \ee

 \be 
\hat\Gamma_{\ln \mathcal{M} \ln \eta}=
 \rho_\rd^2  \,  \frac{1}{60} \left(-72 Q^2+9 S-\frac{18}{2 Q^2+1}+13\right) \, ,
 \ee

 \be 
\hat\Gamma_{\ln \eta \ln \eta}=
 \rho_\rd^2  \, \frac{1}{50} \left(36 Q^2-12 S+\frac{9}{2 Q^2+1}-4\right)
  \, ,
 \ee
 and trivially
 
 \be 
\hat\Gamma_{f_0t_c \, \lambda^i}=
\hat\Gamma_{\phi_c \, \lambda^i}=0
  \, .
 \ee
 
 The analytical approximation and the numerical results  are compared
 in table~\ref{tc} for the initial LIGO detector and two different 
 values of the  total mass.
 
\begin{table}
\caption{Comparison of the analytical approximation and the numerical results,
as described in the text,
for the elements of the ringdown Fisher matrix for the initial LIGO detectors
assuming $a=0.98$ and $\gamma_0=0$.}
\begin{center}
\begin{tabular}{lrr}\\
\hline
\hline
   & analytical & numerical \\
\hline
\multicolumn{3}{l}{ $M = 20 M_\odot$ } \\\hline
$\hat\Gamma_{\ln \mathcal{A}_\ins \ln \mathcal{M}} /  \rho_\rd^2$ & 1.646  &
1.640  \\
$\hat\Gamma_{\ln \mathcal{A}_\ins \ln \eta}/  \rho_\rd^2$ & -0.488  & -0.484  \\
$ \hat\Gamma_{\ln \mathcal{M} \ln \mathcal{M}}/  \rho_\rd^2$  & 288.52  & 276.33  \\
$\hat\Gamma_{\ln \mathcal{M} \ln \eta}/  \rho_\rd^2$ & -172.29  & -164.98  \\
$ \hat\Gamma_{\ln \eta \ln \eta}/  \rho_\rd^2$ & 103.13  & 98.74  \\
\hline
\multicolumn{3}{l}{ $M = 100 M_\odot$ } \\\hline
$\hat\Gamma_{\ln \mathcal{A}_\ins \ln \mathcal{M}}/  \rho_\rd^2$ & 1.299  &
1.276   \\
$\hat\Gamma_{\ln \mathcal{A}_\ins \ln \eta}/  \rho_\rd^2$ & -0.279  & -0.265 \\
$ \hat\Gamma_{\ln \mathcal{M} \ln \mathcal{M}}/  \rho_\rd^2$  & 288.41  & 284.81   \\
$\hat\Gamma_{\ln \mathcal{M} \ln \eta}/  \rho_\rd^2$ & -172.40  & -170.25  \\
$ \hat\Gamma_{\ln \eta \ln \eta}/  \rho_\rd^2$ & 103.30  & 102.02   \\\hline
\end{tabular}
\end{center}
\label{tc}
\end{table}


\end{document}